\documentstyle[preprint,tighten,eqsecnum,epsfig,aps]{revtex}
\input{psfig}

\def\beq{\begin{equation}}   \def\eeq{
\end{equation}}

\begin{document}
\title{Radiation and evolution of small relativistic dipole in QED  } \author{
B. Blok\thanks{E-mail: blok@physics.technion.ac.il} }
\address{Department of Physics, Technion -- Israel Institute of
Technology, Haifa 32000, Israel}
\maketitle

\thispagestyle{empty}

\begin{abstract} We study  in the quasiclassical approximation the
radiation reaction and it's influence on the space-time evolution
for the small relativistic dipole moving in a constant external
electromagnetic field in QED.

 \end{abstract}

 \setcounter{page}{1} \section{Introduction}
\par The problem of the radiation losses (radiation  reaction) of
the particle moving in the given external field is the classical
problem both in quantum and in classical physics that had always
attracted a lot of attention. The classical physics studies are
thoroughly reviewed in refs. \cite{LL,TM,R}.  Since the creation
of QED this problem was thoroughly studied on the quantum level by
several groups of authors, the results reviewed in refs.
\cite{LL1,F,NR,Baier,ST,AS1,AS2}. The two different lines  of
approach were developed, one based on the use of the exact
wave-functions in the external field \cite{F,NR,ST}, and the other
based on the quasi-classical approach \cite{Baier}.
\par In particular, the classical results for the radiation reaction were
extended to the quantum case, and it was shown that  for
ultra-relativistic particles such that $FE/m^3\gg 1$, ($F$ is the
field strength, E $-$ the energy and m the mass of the particle),
the law of radiation reaction changes drastically compared to the
classical case due to the strong recoil effects.
\par Recently a new version of the  quasi-classical approach based on
the use of the quasi-classical Schredinger wave functions was
developed in refs. \cite{AS1,AS2} and references therein.
\par Although the theory of a particle in the external field seems to be
thoroughly developed there is still a lot of interest in the
subject. The reasons, apart from the internal beauty of the
subject include a number of practical reasons. First, the external
field is the simplest model of the media. Second, the QED results
can be viewed as a starting point for the discussion of the
propagation of the QCD particles in the media, the subject being
extremely popular recently due to the recent interest in
quark-gluon plasma \cite{Schifffe}. Next, it was realized that the
space-time evolution of the point charge in the external field is
closely connected with the fundamental properties of QED, leading
to the concept of the semi-bare electron \cite{Feinberg}.
\par The above research was devoted, however, to the radiation reaction of the
charged particle in the external field. Much less is known about
the dipole propagation in the external field. The experimental
research of fast $e^+-e^-$ pairs propagating in the media
continues since fifties, including the famous experiments by
Perkins \cite{Perkins} in 1957. Theoretical investigation of the
fast $e^+-e^-$ pairs leading to the concept of charge
transparency, was started in refs. \cite{Chudakov},
\cite{Iekutieli}. Recently,  there was a renewed theoretical
interest in the study of the relativistic dipole in QED. The
reasons are
 both practical (explanation of the experimental data on $e^+-e^-$ pairs),
and theoretical. In particular, it was realized that the quantum
effects play much bigger role in the propagation of the dipole in
external field than that of the single particle, leading to the
discovery of the quantum diffusion \cite{Farrar}. The essence of
the latter phenomena is the diffusion-type law of the fast dipole
expansion in the weak external field due to the noncoulombic
quantum photon exchange between the components of the dipole. Thus
it was realized, that the study of the propagating QED dipole, and
in particular of it's space-time evolution, is important for the
understanding of the fundamental properties of the QED, in light
of ref. \cite{Feinberg}. Moreover, the study of the propagating
dipole is extremely important due to it's possible generalization
to QCD, where the dipole, due to confinement, may be basic degree
of freedom \cite{FS1,M,FS}. This approach led to the discovery of
color transparency phenomena in QCD \cite{FS1}. Moreover, the QED
dipole is identical to the QCD dipole connected to the
deep-inelastic scattering on the longitudinal virtual photons
\cite{CFS}.
\par However, there is still very few knowledge about the
properties of the propagating relativistic dipole (in particular
relative to what we know about the propagation of the single
charged particles). \par  The main goal of the present paper is to
study the radiation reaction, and in particular the pattern of the
charge transparency, and it's influence on the evolution of the
small ultra-relativistic dipole in the arbitrary strong external
field in QED. In particular we shall be interested in the
influence on the radiation reaction of the interference between
the fields created by different components of the dipole.
 For simplicity we shall consider the case of the
dipole containing two oppositely charged scalar particles of the
same mass, moving in the constant external field whose direction
is orthogonal to the direction of the motion of the center of mass
of the dipole. We shall assume that two particles were created at
the time $T=0$ in the same point of the space-time $\vec r (0)=0$.
\par The main goal of this paper is to take into account the
influence of the quantum effects on the radiation reaction of the
dipole. We shall be able to take into account the quantum effects
connected with the recoil. We will not be able to take into
account the quantum effects connected with the quantum character
of the motion of the dipole, in particular we shall not be able to
take into account the spread of the dipole wave packets and the
quantum diffusion.
  We will not
take into account the spin of the particle, limiting ourselves to
the scalar particle case.
\par Throughout the
paper we   use the quasi-classical wave functions first derived in
refs. \cite{AS1,AS2}.
\par We shall see that there are three distinct time scales:
$1/E\ll T\ll m/F$ (this time regime exists for the dipole such
that the  initial transverse motion of its components is
non-relativistic), $m/F\ll T\ll E/F$, $T\gg E/F$. For the first
regime (we shall call it very small dipole regime) the radiation
reaction is strongly suppressed by interference. The interference
also decreases a number of emitted photons. For the second regime
the decrease in radiation reaction, relative to the sum of
radiation reactions of  two independent particles, depends on the
Lorentz invariant parameter $\chi=\displaystyle{\frac{FE}{m^3}}$.
For $\chi\ll 1$ the interference quickly decreases starting from
$T\sim m/F$. For $\chi\gg 1$ the interference starts to decrease
only starting from the larger time $T^*\sim (E/F^2)^{1/3}$. In the
latter case the dipole effects especially change the photon
spectrum. The relevant photons first are concentrated near the
end-point of the spectrum, and not in the middle, as for a single
charged particle. The maximum of the radiation reaction spectral
curve moves towards saturation at frequencies $\omega \sim 0.4E$.
\par The interference radically changes the frequency
distribution of the number of radiated photons. Instead of
unbounded increase at small frequencies, it now goes to zero as
$\omega\rightarrow 0$ , and has a  maximum at finite frequency, of
the order of the maximum of the radiation reaction.\par Finally in
the third regime, the interference does not influence the
radiation reaction, but still cuts off the soft photons with
$\omega\ll 1/T$, and the photon distributions will have a finite
maximum at $\omega \sim 2/T$.
\par Our results, derived in the approximation of
the constant external field, can be translated to the
model-independent language of the propagation of the dipole
through the arbitrary external media. Indeed, the
Lorentz-invariant parameter $\chi=FE/m^3$ is really a ratio of two
parameters: the parameter $l_c=E/m^2$, which is  (up to
non-important here numerical coefficient ) a coherence length, and
$l_F=m/F$, which is the field regeneration length (or time between
successful interactions with the external field). Thus the
parameter $\chi$ actually measures a number of collisions once the
dipole propagates through the coherence length. In particular
color transparency and quantum diffusion considered in refs.
\cite{Farrar,FS1,FS} correspond to the case $\chi\ll 1$. The
regime of the very small dipole corresponds to the case $T\le
m/F$, i.e., in the model independent language, to the case to the
case when the propagation time is less than a time $T_F$
 one needs to meet an external field photon. In other words, $
 l_F$ is
analogous to the mean free path in the media language. Then it is
clear that in this regime the radiation is always suppressed,
independent of the parameter $\chi$. However, the later time
evolution depends on the parameter $\chi$. If $\chi\ll 1$, and
this corresponds to the case considered in refs.
\cite{Farrar,FS1,FS}, the coherence length is much less than
$l_F$, the radiation suppression ends, as we shall see, at $T\sim
m/F$, and apart from the small time interval in the beginning
$\sim E/m^2\ll m/F$, one can use for the study of the radiation
reaction and the spectra of the emitted photons a quasi-classical
approximation. However, in the opposite case, $\chi \gg 1$, we
have the situation of the multiple collisions during the coherence
length. In this case, we were able to develop a quasi-classical
theory of the radiation emission taking into account recoil. Our
results show the suppression of the radiation reaction and the
photon emission up to the time $T^*\gg T_F$. This looks quite
similar to the Landau-Pomeranchuk effect for the propagation of
the fast particle in the media. There the effect also appears when
the coherence length is much bigger than the free mean path
\cite{AS1}. However, as we discuss below, the classical
approximation may be not applicable to the situation when $l_F\ll
l_c$ for the dipole. This is the case that occurs in the
statistical mechanics, when the coherence length is bigger than
free mean path.Then there is a number of the important effects
that arise only beyond the quasi-classical approximation
\cite{Imri}.  In this paper we shall only study what one obtains
sticking to the quasi-classical approximation. The results may be
considered as a starting point for the future study.
 \par The paper is
organized in the following way. In Chapter 2 we shall consider
small classical dipole, but will derive it's radiation reaction
using relativistic quantum mechanics, and check that the classical
approach  corresponds to the recoiless limit of the
quasi-classical theory. We shall review the results for a single
particle, then consider the case of the radiation of the arbitrary
dipole, and then derive the radiation reaction in the small and
very small dipole limits. In Chapter 3 we shall briefly review the
classical wave-functions method of refs. \cite{AS1,AS2} and extend
it to the case of the arbitrary dipole. Next we shall assume that
the dipole is small (in the plane transverse to the direction of
it's center of mass motion) and derive the general formula for the
radiation reaction of such small dipole. In the sections 4 and 5
we shall use the above formulae to study the radiation reaction in
two important limiting cases. In Chapter 4 we shall study the
frequency distribution of radiation and the time dependence of the
total radiated energy for the limit of very small times, when the
dipole own field was not generated yet. We shall call this regime
a very small dipole regime. This regime can be also characterized
as the regime when the particle deflection angle due to external
field is less that the radiation angle. In Chapter 5 we shall
consider the scale of times when the dipole is still small, but
it's field has already been generated. The particle deflection
angle is much bigger than the radiation angle. We shall study the
frequency distribution of the photons and the radiation reaction
also in this case. We shall see that the radiation reaction
depends on the parameter $\chi=FE/m^3$. (Recall that for the
single fast moving charged particle radiation reaction
qualitatively depends on this parameter that is a Lorentz
invariant: $\chi=\sqrt{(F_{\mu\nu}p^{\nu})^2/m^6}$ \cite{LL}.) In
Chapter 6 we shall study the total back-force acting on the dipole
for very small times and it's influence on both the transverse and
the longitudinal evolution of the dipole. In Chapter 7 we shall
make some qualitative comments on the influence of the quantum
nature of the dipole motion on the radiation reaction, in
particular on the possibility to go beyond the quasi-classical
approximation.
 Our results, the
directions for the future work and possible implications for QCD
will be summarized in the conclusion.
\section{Radiation Reaction of the Fast Relativistic Dipole.}
\subsection{Radiation of the single scalar particle}
\par Let us start by briefly recalling the basic quasi-classical formalism
for radiation of photons by relativistic charged particle without
taking into account recoil \cite{LL1}. The results are the same as
obtained by using classical electromagnetism theory \cite{LL,AS1},
but we shall use from the beginning not the wave but the photon
formalism, that will be easily extended in the next chapter to the
case when we need to take into account recoil and the classical
electromagnetism theory  will be unapplicable.
\par The matrix element of the
interaction between the electromagnetic field and the scalar
particle is given by \beq S^{(1)}=-iq\int d^4x
A_{\mu}(x)J^{\mu}(x), \label{111} \eeq where $J_{\mu}(x)$ is the
current density operator in the external field, \beq
J^{\mu}=\Phi^*(P^{\mu}\Phi)-(P^{\mu}\Phi^*)\Phi . \label{112} \eeq
The operator $P^{\mu}$ is the generalized momentum operator in the
external field. Consequently, the matrix element for the
 emission of the photon with the frequency $\omega$, wave vector
$\vec k$ and polarization vector $\vec e$ is given by
 \begin{equation}
M_{\rm fi}=-iq\int^T_0dt\int d^3
\vec r \sqrt{\frac{2\pi}{\omega}}\frac{1}{\sqrt{E_iE_f}}\phi^*_f(\vec r,t)
 (\vec e\cdot \hat{P})\exp{(i(\omega t-\vec k\cdot r)}\phi_i(\vec r,t)
\label{1} \end{equation} Here $\phi_i$ is the initial and $\phi_f$
is the final state wave functions, normalized by the condition
\beq \int d^3\vec r\phi^*(\vec r)\phi (\vec r)=1. \label{113} \eeq
 The operator $\vec P$ is
 $$\vec P =\vec p - q\vec A,$$
$\vec p=-\frac{\partial}{\partial x_i}$ is the momentum , q is the
charge of the particle and $A(\vec r ,t)$ is the vector potential.
$E_i$ and $E_f$ are the energies of the initial and the final
states.
\par We shall use the quasi-classical wave functions of the scalar
particle in the external field: \beq \phi (\vec r,
t)=\sqrt{\frac{D}{E_i-qA_0}}\exp{(\frac{i}{\hbar}S(r,p,t))}.
\label{q} \eeq Here $S(r,p,t)$ is the action of the particle with
the momentum $\vec p$
  calculated along the classical trajectory of the particle
passing through the point with the coordinate $\vec r$ at the time
t and having the momentum $\vec p$ at t=0.
 D is
the Van-Vleck determinant: \beq D=\sqrt{\Vert\frac{\partial^2
S(\vec r, \vec p)}{\partial \vec r\partial \vec p}\Vert}=
\frac{1}{E}\sqrt{\delta (\vec r-\vec r(t))}. \label{D} \eeq The
wave functions (\ref{q}) can't be substituted directly into the
matrix element (\ref{1}), due to the appearance of the quickly
oscillating factors $$\exp{i(S(r,p_f,t)-S(r,p_i,t))/\hbar } $$ for
$\hbar\rightarrow 0$. In order to avoid this difficulty we have to
use the representation:
\begin{equation}
\phi_i(\vec r ,t)=\int d^3\vec p\phi_p(\vec r,t)S_{\rm pp_0},
\label{2}
\end{equation}
where $\phi_p$ is the quasi-classical wave function of the
particle in the external field possessing at $t\rightarrow \infty$
the asymptotics
$$\phi_p(\vec r,t)\rightarrow \frac{1}{\sqrt{2E_p}}\exp{(i (\vec p\vec r-Et))}).$$
$S_{\rm pp_0}$ is the scattering matrix of the particle in the
external field considered.  If we neglect the recoil , and
substitute the representation (\ref{2}) for the final state wave
function into the matrix element (\ref{1}), we shall recover the
classical amplitude for the radiation of the electromagnetic
waves, and the classical expression for the energy loss during a
time interval  T. (see e.g. refs. \cite{Baier,AS1} for details):
 \begin{eqnarray}
dW_{\rm cl}&=&\displaystyle{\frac{2q^2}{\pi^2}}(d^3k )
\int^T_0\int^T_0dtdt'(\vec e\cdot \vec
v(t)(\vec e^*\cdot \vec v(t')\nonumber\\[10pt]
&\times &\exp{(i\omega (t-t')-i\vec k\cdot (\vec r(t)-\vec r(t'))}.\nonumber\\
\label{5}
\end{eqnarray}
After averaging over the photon polarization vectors we obtain
 \begin{eqnarray}
dW_{\rm cl}&=&q^2\displaystyle{\frac{1}{2\pi^2}d^3\vec k}
\int^T_0\int^T_0dtdt'(\vec v(t)\cdot \vec v(t')-(\vec n\cdot \vec
v(t))(\vec n\cdot \vec
v(t'))\nonumber\\[10pt]
&\times&\exp(i\omega (t-t')-i\omega\vec n\cdot (\vec r(t)-\vec r(t'))\nonumber\\
\label{500}
\end{eqnarray}
\par
where $\vec k=\omega \vec n$.
 Note  that
 $\vec n\cdot \vec v(t)=\vec v\nabla_{\vec r} =\partial/\partial t$. Hence
the terms in the latter equation containing $\vec n$ in the
preexponential can be integrated by parts:
 \begin{eqnarray}
dW_{\rm fi}&=&q^2\displaystyle{\frac{1}{2\pi^2}d^3k}
(\int^T_0\int^T_0dtdt'(\vec v(t)\cdot \vec v(t')-1) \times
\exp(i\omega (t-t')-i\omega\vec n\cdot (\vec r(t)-\vec
r(t'))\nonumber\\[10pt]
&+&\displaystyle{\frac{4}{\omega}}\int^T_0\sin{(\omega T+\vec
n\vec r(T))/2)}\cos{(\omega (T-2s)+\vec n\cdot(\vec r(T)-2\vec r
(s)))/2}\nonumber\\[10pt]
&-&\displaystyle{\frac{2}{\omega^2}}(1-\cos{\omega T+\omega\vec
n\cdot\vec r(T)}). \label{501}
\end{eqnarray}
It is straightforward to see however, that the last two lines in
eq. (\ref{501}) correspond to terms decreasing or bounded with T,
while the expression in the first line increases with $T$. Thus
the two last lines can be omitted if we are interested in large
time intervals $T\gg 1/\omega$. Indeed, we can integrate eq.
(\ref{501}) over the photon direction $\vec n$ and obtain
 \begin{eqnarray}
dW_{\rm cl}&=&q^2\displaystyle{\frac{2}{\pi}\omega d\omega }
(\int^T_0\int^T_0dtdt'(\vec v(t)\cdot \vec v(t')-1) \times
\cos{(\omega (t-t')}\displaystyle{\frac{\sin{(\omega\vert\vec
r(t)-\vec r(t')\vert)}}{\vert\vec r(t)-\vec r(t')\vert}} \nonumber\\[10pt]
&+&\displaystyle{\frac{2}{\omega}\int^T_0\frac{\cos{(\omega s-
\omega r(s))}-\cos{(\omega s+\omega r(s))
}}{r(s)}}\nonumber\\[10pt]
&+&\displaystyle{\frac{2}{\omega}\int^T_0\frac{\cos{(\omega (T-s)-
\omega\vert\vec r(T)-\vec r(s)\vert )}-\cos{(\omega s+\omega
\vert\vec r(T)-\vec r(s)\vert )
}}{\vert\vec r(T)-\vec r(s)\vert }}\nonumber\\[10pt]
&-&\displaystyle{\frac{2}{\omega^2}(1-\frac{\sin{\omega
(T+r(T))}-\sin{(\omega T-r(T))}}{r(T)}}.\nonumber\\ \label{503}
\end{eqnarray}
It is easy to see that the last 3 lines in the eq. (\ref{503}) are
suppressed like $1/(\omega T)$ relative to the double integral in
the first line, and thus can safely discarded if we are interested
in the frequencies and time intervals $\omega T\gg 1$. In order to
know numerically how big are these terms, we shall however keep
them.
\par Finally, since we are usually interested in the energy losses
in the unit of time, we can differentiate eq. (\ref{503}) over
time T and obtain
 \begin{eqnarray}
\displaystyle{\frac{dW_{\rm cl}}{dT}}
&=&q^2\displaystyle{\frac{4}{\pi}\omega d\omega } (\int^T_0dt(\vec
v(T)\cdot \vec v(t)-1) \times \cos{(\omega
(T-t)}\displaystyle{\frac{\sin{(\omega\vert\vec
r(T)-\vec r(t)\vert)}}{\vert\vec r(T)-\vec r(t)\vert}} \nonumber\\[10pt]
&+&\displaystyle{\frac{2}{\omega}\frac{\cos{(\omega T- \omega
r(T))}-\cos{(\omega T+\omega r(T))
}}{r(T)}}\nonumber\\[10pt]
&+&\displaystyle{\frac{2}{\omega}\int^T_0\frac{d}{\frac dT}
\frac{\cos{(\omega s)- \omega\vert\vec r(T)-\vec r(T-s)\vert
)}-\cos{(\omega s+\omega \vert\vec r(T)-\vec r(T-s)\vert )
}}{\vert\vec r(T)-\vec r(T-s)\vert }}\nonumber\\[10pt]
&-&\displaystyle{\frac{d}{dT}\frac{2}{\omega^2}(1-\frac{\sin{\omega
(T+r(T))}-\sin{(\omega T-r(T))}}{r(T)}}. \label{504}
\end{eqnarray}
\par Below we shall use  the first line in the latter
equation and check that the last 3 lines can be neglected:
 \beq
\displaystyle{\frac{dW}{dT}} =q^2\displaystyle{\frac{4}{\pi}\omega
d\omega } (\int^T_0dt(\vec v(T)\cdot \vec v(t)-1) \times
\cos{(\omega (T-t)}\displaystyle{\frac{\sin{(\omega\vert\vec
r(T)-\vec r(t)\vert)}}{\vert\vec r(T)-\vec r(t)\vert}}.
\label{505} \eeq The latter equation, if the limits of integration
are infinite, can be easily brought into the standard form of the
classical electromagnetic theory \cite{LL1,AS1}.
\subsection{Radiation of the relativistic dipole: general theory.}
\par Let us now consider the radiation reaction of the relativistic dipole
in the case we can neglect recoil, i.e. $\omega \ll E$.
\par
 For simplicity we  consider the symmetric dipole, whose center of mass
moves with the speed $v\sim c$ in the direction orthogonal to the
direction of the constant external field, and which was created at
time $T=0$ in the point $\vec r (0)=0$. We shall denote the
components of the dipole as P and A (particle and antiparticle).
Let us assume that the particles of the dipole have, after it's
creation, the same initial energy $E_i$, and the orthogonal
component of the velocity $v_{\rm 0t}$. Note that if $u_{0t}$ is
the velocity in the transverse plane in the center of mass
reference frame, moving with dipole, then
$v_{0t}=\frac{m}{E}u_{0t}$, meaning that in any case $v_{0t}\le
m/E$. In our kinematics the two components of the dipole will have
the same velocity component in the direction of c.m. motion and
the opposite sign components in the transverse plane.
 \par The amplitude of the radiation of the photon with the polarization vector $\vec e$
the wave vector $\vec k$ and the frequency $\omega$
 will be the difference
(due to the different charges of the dipole components) of the
amplitudes of the photon emission of the particle and the
anti-particle components of the dipole. Using the equations of the
previous chapter it is straightforward to write:
\begin{eqnarray}
 M_{\rm fi}=-&i&\displaystyle{\sqrt{\frac{2\pi}{\omega}}\frac{1}{sqrt{E_iE_f}}}\int^T_0dt(\vec
e\cdot \vec v_P (t)
\exp{i(\omega t-\vec k\cdot \vec r_P(t))}\nonumber\\[10pt]
&-&\vec e\cdot \vec v_A (t)\exp{i(\omega t-\vec k\cdot\vec
r_A(t)))}.\nonumber\\[10pt]
\label{6}
\end{eqnarray}
Here $\vec r_P(t)$ and $\vec r_A(t)$ are the radius vectors of the
particle and antiparticle components of the dipole. The energy
radiation loss during the time from the creation of the dipole at
the time  $t=0$ till time equal T with the photons radiated in the
frequency range $d\omega $ and the solid angle range $do$ is
\begin{eqnarray}
dW=&&\displaystyle{\frac{q^2}{4\pi^2}}\omega^2 do d\omega
\int^T_0dt\int^T_0dt'
 \exp{i\omega (t-t')}(\vec e\cdot \vec v_P(t)
\exp{i (\vec k\cdot r_P(t)}-\nonumber\\[10pt]
&-&\vec e\cdot \vec v_A(t)\exp{i(\vec k\cdot\vec r_A(t)})(\vec
e^*\cdot \vec v_P(t')
\exp{-i (\vec k\cdot r_P(t')}-\vec e^*\cdot \vec v_A(t')\exp{-i(\vec k\cdot \vec r_A(t')}).\nonumber\\
\label{7}
\end{eqnarray}
 Summing over the polarizations of the photon we obtain
\begin{eqnarray}
\displaystyle{\frac{dW}{d\omega
}}&=&\displaystyle{\frac{q^2}{4\pi^2}}\omega^2
\int^T_0dt\int^T_0dt'
(\exp{i(\omega (t-t'))})\nonumber\\[10pt]
 (v_P(t&)&\cdot v_P(t')-(\vec n\cdot\vec v_P(t))(\vec n
\cdot\vec v_P(t'))
 \cdot \exp{i(\vec k\cdot(\vec r_P (t)-\vec r_P
(t'))} +(P\leftrightarrow
A)-\nonumber\\[10pt]
 &-&((\vec v_P(t)\cdot\vec v_A(t')-(\vec n\cdot\vec
v_P(t))(\vec
n\cdot\vec v_A(t'))\exp{i(\vec k\cdot (\vec r_P (t)-\vec r_A (t'))} +(P\leftrightarrow A)).\nonumber\\
\label{8}
\end{eqnarray}
Here $\vec k=\omega \vec n$. Using, as for the single particle,
 $\vec n\cdot v(t)=\vec v\nabla_{\vec r} =\partial/\partial t$, we can carry the integration by parts
 and obtain:
\begin{eqnarray}
\displaystyle{\frac{dW}{d \omega do}}
&=&\displaystyle{\frac{q^2}{4\pi^2}}
\omega\int^T_0dt\int^T_0dt'\exp{i(\omega (t-t'))} (\vec
v_P(t)\cdot \vec v_P(t')-1)\exp{i\vec k\cdot (\vec r_P (t)-\vec
r_P (t'))} +(P\leftrightarrow
A)\nonumber\\[10pt]
&-&(\vec v_P(t)\cdot \vec v_A(t')-1)\exp{i\vec k\cdot(\vec r_P
(t)-\vec r_A (t'))}
+(P\leftrightarrow A)\nonumber\\[10pt]
&+&\Delta W (\omega , T).\nonumber\\
 \label{10}
\end{eqnarray}
\par The latter equation gives us the formula for the radiation of the
arbitrary relativistic dipole. Note that  it is a sum of two terms
that correspond to the radiation of the single particle and two
terms that correspond to the interference between the particle and
the antiparticle.
\par The term $\Delta W$ arises from the integration by parts (cf. the single particle) and
is equal to
\begin{eqnarray}
\Delta W(\omega,T)&=&do\displaystyle{\frac{q^2}{4\pi^2}}\omega^2
(\displaystyle{\frac{2}{\omega}}\int^T_0\sin{(\omega
(T-s)+\omega\vec n(\vec r_P(T)-\vec r_P(s))}+(P\leftrightarrow
A)\nonumber\\[10pt]
&-&\displaystyle{\frac{2}{\omega}}\int^T_0 ds (\sin{(\omega
(T-s)+\omega\vec n(\vec r_P(T)-\vec r_A(s))}+\sin{(\omega
(T-s)+\omega\vec n(\vec r_A(T)-\vec r_P(s))}+\nonumber\\[10pt]
&-&(P\leftrightarrow A)-
\displaystyle{\frac{2}{\omega^2}}(1-\cos{\omega\vec n\cdot (\vec
r_P(T)-\vec r_A(T)})). \label{142}
\end{eqnarray}
\par We can  integrate over the angle variable $do$ and obtain:
\begin{eqnarray}
\displaystyle{\frac{dW}{d \omega }}
&=&\displaystyle{\frac{q^2}{\pi}}\int^T_0dt\int^T_0dt'\cos{(\omega
(t-t'))} (\vec v_P(t)\cdot \vec
v_P(t')-1)\displaystyle{\frac{\sin{\vert \vec r_P (t)-\vec r_P
(t'))\vert}}{\vert\vec r_P(t)-\vec r_P(t')\vert )}}
+(P\leftrightarrow
A)\nonumber\\[10pt]
&-&(\vec v_P(t)\cdot \vec v_A(t')-1)\displaystyle{\frac{\sin{\vert
\vec r_P (t)-\vec r_A (t'))\vert}}{\vert\vec r_P(t)-\vec
r_A(t')\vert )}}
-(P\leftrightarrow A)\nonumber\\[10pt]
&+&\Delta G (\omega , T).\nonumber\\
\label{643}
\end{eqnarray}
Here the term $\Delta G$ corresponds to the integral of the
$\Delta W$:
\begin{eqnarray}
\delta
G&=&\displaystyle{\frac{2q^2\omega}{\pi}}(\int^T_0ds(\displaystyle{\frac{\cos{(\omega
(T-s)-\omega\vert \vec r_P(T)-\vec r_P(s)\vert)}}{\omega\vert \vec
r_P(T)-\vec r_P(s)\vert)}}-\displaystyle{\frac{\cos{(\omega
(T-s)+\omega\vert \vec r_P(T)-\vec r_P(s)\vert)}}{\omega\vert \vec
r_P(T)-\vec r_P(s)\vert)})}\nonumber\\[10pt]
&+&(A\leftrightarrow P)\nonumber\\[10pt]
&-&(\displaystyle{\frac{\cos{(\omega (T-s)-\omega\vert \vec
r_P(T)-\vec r_A(s)\vert)}}{\omega\vert \vec r_P(T)-\vec
r_A(s)\vert)}}-\displaystyle{\frac{\cos{(\omega (T-s)+\omega\vert
\vec r_P(T)-\vec r_A(s)\vert)}}{\omega\vert \vec
r_P(T)-\vec r_A(s)\vert)})}-(A\leftrightarrow P)\nonumber\\[10pt]
&-&\displaystyle{\frac{1}{\omega}(1-\frac{\sin{\omega\vert \vec
r_P(T)-\vec r_A(s)\vert)}}{\omega\vert \vec r_P(T)-\vec
r_A(s)\vert)})}.\nonumber\\
 \label{642}
\end{eqnarray}
In order to get the radiation reaction it is enough to
differentiate the above equations over T. \par The latter
equations describe the radiation of the arbitrary relativistic
dipole, integrated over the angles, for the time interval $T$.
\par Now we can move to our goal $-$ to consider the case of the small relativistic dipole.
\subsection{Radiation of the small relativistic dipole.}
\par Consider small quasiclassical relativistic dipole, i.e. $v\gg v_t$, where
$v$ is it's center of mass velocity and $v_t$ is the transverse
component of the velocity ($v_t$ can be both relativistic and
nonrelativistic). For sufficiently small times one can estimate
\beq v_t (T)\sim v_{0t}+FT/E. \label{301} \eeq Here F is the
external field: \beq \vec F=\vec E+\vec v\times \vec
H,\label{F}\eeq $\vec E$ is an electric and $\vec H$ is a magnetic
field.
 Consequently, one considers dipole as small if
\beq FT\ll E \label{302}. \eeq For bigger time scales,
$$FT\ge E,$$
 the components of the dipole  behave as
independent particles and there is no interference.
\par Let us study the interference pattern in the small dipole.
\par Let us  assume that the condition (\ref{302}) is satisfied.
 Then the photons are radiated into the small cone around z axis (we choose the z
axis in the direction of the propagation of the dipole), of order
$m/E$ at $T\sim 1/E_i$, where $E_i$ is the initial energy of each
of the components of the dipole. Later the radiated photons are
concentrated in two cones round the directions of the components
of the dipole. It is clear that there exist, even if the condition
(\ref{302})) is satisfied, two distinct possibilities: the two
radiation cones, generated by the dipole components overlap, and
that they stop to overlap. Since the cone angle for the
ultra-relativistic particle is $\theta\sim m/E_i$, we see that the
condition that the cones overlap is \beq v_{ot}+FT/E\le m/E
\label{501A}
 \eeq If
we can neglect the initial transverse velocity, the latter
condition becomes
 \beq
  T\le m/F\label{502}
\eeq If
$$m/F\ll T\ll E/F$$
the dipole is still small, but the cones do not overlap, and the
interference must decrease drastically. There is also the
self-consistency condition: since $T\gg 1/E$, we must have \beq
mE\gg F\label{PC1}\eeq for the possibility of considering the very
small dipole, with the cones overlapping, quasi-classicaly.We need
the weaker condition \beq E^2\gg F,\label{PC2}\eeq for the
possibility to consider small quasi-classical dipole. If the
latter conditions are not fulfilled, we must take into account the
interference of the dressing by external field and the generation
of the self-field by bare particle.This is beyond the scope of
this research. (Although it can be that our analysis is
qualitatively true even in the latter case, since the self
dressing generates usually quickly oscillating terms that can be
singled out).
\par We conclude that the classical dipole has two regimes:
1) very small dipole, when the radiation cones of the particle and
the anti-particle overlap strongly, $T\ll m/F$, and 2) small
dipole in the sense it still moves along z axis, but the cones of
the radiation do not overlap, and the interference decreases. Note
that these two cases  correspond to two possible relations between
the depletion angle of the single charged particle in the external
field and the radiation angle. The very small dipole corresponds
to the case when the latter angle is much bigger than the former
and the small dipole $-$ when the former is bigger than the
latter. Note also that for a relativistic in the c.m. of dipole
transverse motion means  (then $v_t\sim m/E$) we have only small
dipole regime.
\par Suppose we have the very small dipole. Let us analyze the
interference pattern.
 Consider
 the exponents in eq. (\ref{10}).
The exponents in the terms that contain only the particle or only
the anti-particle radiation are
$$\omega (\cos{\theta}( z(T)-z(t))+\sin{\theta} (y(T)-y(t)).$$
Here $\theta$ is the angle between the photon wave vector and
direction of the z axis, and
$$z(T)-z(t)\sim v_z (T-t), y(T)-y(t)\sim v_t (T-t)$$
It is clear that the corresponding integrals will be saturated by
$t\sim T$, and the first term will be dominant since
$\sin{\theta}\ll 1$ Consider now the exponents in the interference
terms in eq. (\ref{10}). These exponents have the form, for the
chosen kinematics
$$\omega (\cos{\theta}( z(T)-z(t))+\sin{\theta}\sin{(\phi )} (y(t)+y(T)).$$
Here $\phi$ is the asimutal angle. In the first approximation we
can put $y(t)\sim y(T)=d(T)/2$ in the latter equation, and instead
of integrating, substitute $\sin{\theta}$ by it's characteristic
value $m/E$. (d(T) is the scale of the dipole, i.e. the separation
between the charges, which in our kinematics is purely
transverse). Then integral over the angle $\phi$ gives the Bessel
function: \beq \frac{1}{2\pi}\int^{2\pi}_0\exp{(i\sin{\phi}\omega
d(T)m/E)}d\phi= J_0(\displaystyle{\frac{\omega}{E}}d(T)m)\label{J}
\eeq With the same accuracy we can substitute $\cos\theta
(z(t)-z(T))$ with $\cos\theta (z(t)-z(T))+\sin\theta
\sin{\phi}(y(t)-y(T)$, i.e. after taking into account the
interference term, the exponent in the interference term will be
the same as in the direct terms. Then for very small dipole we can
rewrite eq. \ref{10} as
\begin{eqnarray}
\displaystyle{\frac{dW}{d \omega dT}}
&=&\displaystyle{\frac{4q^2}{\pi}}\int^T_0dt\cos{(\omega (T-t))}
(\vec v_P(T)\cdot \vec v_P(t)-1)\displaystyle{\frac{\sin{\vert
\vec r_P (T)-\vec r_P (t))\vert}}{\vert\vec r_P(T)-\vec
r_P(t)\vert )}}
\nonumber\\[10pt]
&\times&(1-J_0(\displaystyle{\frac{\omega}{E}}md(T))\nonumber\\
\label{644}
\end{eqnarray}
In addition, there is contribution from the terms, that correspond
to integration by parts, where it is enough to do the same
approximation:
\begin{eqnarray}
\displaystyle{\frac{dG}{dTd\omega}}&=&\displaystyle{\frac{q^2}{\pi}}
\displaystyle{\frac{2}{\omega}\int^T_0\frac{d}{dT}
\frac{\cos{(\omega s)- \omega\vert\vec r(T)-\vec r(T-s)\vert
)}-\cos{(\omega s+\omega \vert\vec r(T)-\vec r(T-s)\vert )
}}{\vert\vec r(T)-\vec r(T-s)\vert }}\nonumber\\[10pt]
&\times&(1-J_0(\displaystyle{\frac{\omega}{E}}md(T))\nonumber\\[10pt]
&+&\displaystyle{\frac{d}{dT}\frac{2}{\omega^2}\frac{\sin{\omega
d(T)}}{T}}.\nonumber\\
\label{ada}
\end{eqnarray}
 Here $ \dot{\vec d}(s)$ is the time derivative of
the dipole moment, i.e. the relative velocity of the particle and
antiparticle:
\begin{equation}
 \dot{\vec d}(s)=\frac{\partial (\vec r_P(s)-\vec
r_A(s))}{\partial s}. \label{12}
\end{equation}
 The latter equations gives the  radiation energy losses  rate for the very small
relativistic dipole between times $0 $ and T, emitted in the
particular interval of photon frequencies.
\par Note that our interference analysis could be made in terms
not of the characteristic radiation angles, but in terms of the
longitudinal and transverse momenta. Our characteristic angles
$m/E$ correspond to the characteristic transverse momentum of the
emitted photons $q_t\sim m\omega /E$, in particular, if we
consider photons, whose energy is a finite part of E, the
characteristic transverse momentum will be $q_t\sim m$.
\par Consider now the next regime, $E/F\gg T\gg m/F$. This is the
case of the small, but not very small dipole. In this case we can
still consider the trajectory of each of the particles as the
almost straight line. We can follow the above derivation of the
interference terms, but in this case, although still $\theta\ll
1$, we need to take as $\theta$ the angle $v_t/v\sim v_t\sim
v_{0t}+FT/E$. We then get the equation similar to eq. (\ref{644}),
but with the different argument for the Bessel Function:
\begin{eqnarray} \displaystyle{\frac{dW}{d \omega dT}}
&=&\displaystyle{\frac{4q^2}{\pi}}\int^T_0dt\cos{(\omega (T-t))}
(\vec v_P(T)\cdot \vec v_P(t)-1)\displaystyle{\frac{\sin{\vert
\vec r_P (T)-\vec r_P (t))\vert}}{\vert\vec r_P(T)-\vec
r_P(t)\vert )}}\nonumber\\[10pt]
&\times&(1-J_0(\omega \theta (T)d(T))),\nonumber\\
\label{647}\end{eqnarray} where \beq \theta
(T)=v_{0t}+FT/E=v_y(T).\label{900}\eeq Since for the classical
dipole $d(T)\sim v_y(T)T\sim v_{ot}T+FT^2/(2E)$, we see that
interference is suppressed as
$1-J_0(\displaystyle{\frac{\omega}{E}}(v_{0t}^2TE+3v_{ot}FT^2/2+F^2T^3/(2E^2))$
Since $v_t\le m/E$ (due to the relativistic law of the velocity
summation), and $T\gg m/F$, the third term in the argument of the
Bessel function will be dominant, i.e. the interference decreases
as $J_0((\omega/E)(F^2T^3/E))$, and quickly becomes negligible.
Finally note that for very small frequencies one always has
interference. \par Note also that the argument of the Bessel
function can be represented as
$$xb(\tau),\,\,\, b(\tau)=m\frac{d d^2(\tau )}{d\tau}$$
i.e. as a Lorentz invariant (see also the discussion below). Here
$\tau$ is the proper time in the reference frame of the c.m. of
the dipole.
\par Finally, since the integrands in eqs. (\ref{644}) and
(\ref{647}) are concentrated near $T=s$, we can expand them in
Taylor series near $s=T$. Consider first the difference $\vec r
(T)-\vec r(s)$ in the argument of the exponents.
\par For small
dipole it is possible to use the approximations \cite{AS1,AS2}:
\beq \vec v(T)=v(0)(1-v^2_t(T)/2v(0)^2)+\vec v_t (T) \label{as}
\eeq and \beq  \dot{\vec v}(T)_t=q\vec F/E.\label{drot}\eeq
 In this
approximation up to the terms of order $m/E$
$$\dot{\vec v}(T)\sim \dot{\vec v}_t\sim q\vec F/E,$$
i.e. the vectors $\vec v $ and $d\vec v/dT$ are orthogonal. Also
$$\frac{d^2\vec v}{dT^2}=-\omega_0^2\vec v(T),
$$
$$\omega_0=qF/E.$$
\par Then
\begin{eqnarray} \vert \vec r(T)-\vec r (s)\vert
&=&\sqrt{v^2(T)(T-s)^2(1-\omega_0^2(T-s)^2)^2+\omega_0^2(T-s)^2/4}\nonumber\\[10pt]
&\sim & v(T)(T-s)(1+\omega_0^2(T-s)^2/24).\nonumber\\
\label{ad}
\end{eqnarray}
\par Then we  have in the standard way (\cite{AS1}):
\beq \vec v_P(T)\vec
v_P(t)-1=-(1-v^2(T)+(T-t)^2\omega^2_0/2)\label{den} \eeq In the
same approximation \beq \dot{\vec d}(T)\cdot \dot{\vec
d}(s)=4(v_t(T)^2-(T-s)\omega_0v_t(T))\label{final}\eeq Thus we
have our final result for the radiation of the small dipole:
\begin{eqnarray} \displaystyle{\frac{dW}{d \omega dT}}
&=&-\omega\displaystyle{\frac{4q^2}{\pi}}\int^T_0dt
\displaystyle{\frac{(1-v^2(T)+\omega_0^2(T-t)^2/2}{T-t}}
\sin{(\omega (1-v)(T-t))+\omega_0^2(T-t)^3/24)}\nonumber\\[10pt]
&\times&(1-J_0(\omega \theta (T)d(T))).\nonumber\\ \label{681}
\end{eqnarray}
 Here \begin{eqnarray} \theta (T)=m/E\,\,\,\, T\ll
m/F\nonumber\\[10pt]
\theta (T)\sim v_t(T)\sim m/E+FT/E, \,\,\, E/F\gg T\gg
m/F.\nonumber\\
\label{il}\end{eqnarray} In order to obtain the full radiation
reaction we must integrate the latter formulae over $\omega$. We
thus obtained classical radiation reaction of the dipole using
simple wave mechanics. In parallel we understood the nature of the
interference in the transverse plane, that we shall use in the
quantum case.
\section{Radiation reaction for the relativistic dipole: recoil
effects}
\subsection{Single Particle.}
\par In the previous section we studied, using the relativistic quantum mechanics method, the classical radiation from the
classical dipole. Let us now move to the quantum effects. There
are two types of quantum effects \cite{AS1}: first, the effects
due to the quantum character of the particle motion in the
external field. This effect is characterized by the parameter
$F/E^2$ \cite{AS1}. Second there are quantum effects, specifically
due to the motion of the quantum dipole \cite{Farrar,FS1,FS}.
These effects we will not take into account. Third, there are the
recoil effects, that arise if we take into account $E_i\ne E_f$.
The general theory of such effects was first derived in ref.
\cite{Baier}. Recently a new approach was derived by Akhiezer and
Shulga \cite{AS1,AS2}. Let us briefly review the idea of ref.
\cite{AS1}. We return to the derivation of matrix element of the
radiation of photon (\ref{1}). We still use the representation
(\ref{2}) for the quasi-classical wave functions, but when we
substitute them into the matrix element (\ref{1}) we take into
account that the corresponding integral over $\vec p$ is saturated
not at $\vec p=\vec p_f$, as we assumed when we neglected recoil,
but at $\vec p\sim \vec p_f+\vec k$, where as usual $\vec k$ is
the wave vector of the emitted photon. Then it is possible to
prove that the generalized action $S=S_f-(\omega t-\vec k\cdot\vec
r)$ satisfies the generalized Hamilton-Jacobi equation \beq
\displaystyle{\frac{\partial S}{\partial t}}=(\vec \nabla S-q\vec
A-\vec k)^2+m^2 .\label{HY} \eeq Solving this equation and
substituting the solution into the matrix element (\ref{1}), where
we use for the wave function $\phi_i$ the representation
(\ref{2}), one obtains the quasi-classical matrix element of the
photon radiation where the recoil is taken into account:
 \begin{equation}
M_{\rm fi}=-iq\int^T_0dt\int d^3 \vec r
\sqrt{\frac{2\pi}{\omega}}\sqrt{\frac{E_i}{E_f}}(\vec e\cdot\vec
v(t)
 \exp{(i\frac{E_i}{E_f}(\omega t-\vec k\cdot r(t))}.
\label{1111} \eeq Here we can put $E_f=E_i-\omega $.
\par The corresponding radiation reaction will be
 the same as for the single particle in the previous section, except
the rescaling of the frequency $$\omega \rightarrow \omega \frac{E_i}{E_f}$$ in the exponent and the general
multiplier $E_i/E_f$:
\begin{eqnarray}
dW_{\rm fi}&=&q^2\displaystyle{\frac{2}{\pi^2}}\frac{E}{E_f}d^3k
\int^T_0\int^T_0dtdt'(\vec e\cdot \vec
v(t)(\vec e^*\cdot \vec v(t')\nonumber\\[10pt]
&\times &\exp{(\frac{E}{E_f})(i\omega (t-t')-i\vec k\cdot (\vec r(t)-\vec r(t'))}.\nonumber\\
\label{51}
\end{eqnarray}
After averaging over the photon polarisations and integrating by
parts we obtain, as in the previous section, the equation for the
radiation reaction of the single particle including the recoil
effects:
 \begin{eqnarray}
\displaystyle{\frac{dW}{dT}} &=&\omega
q^2\displaystyle{\frac{4}{\pi}d\omega }
\displaystyle{(\int^T_0dt(\vec v(T)\cdot \vec v(t)-1) \times
\cos{(\frac{E}{E_f}\omega
(T-t)}}\displaystyle{\frac{\sin{(\frac{E}{E_f}\omega\vert\vec
r(T)-\vec r(t)\vert)}}{\vert\vec r(T)-\vec r(t)\vert}} \nonumber\\[10pt]
&+&\displaystyle{\frac{2}{\frac{E}{E_f}\omega}\frac{\cos{\frac{E}{E_f}(\omega
T- \omega r(T))}-\cos{\frac{E}{E_f}(\omega T+\omega r(T))
}}{r(T)}}\nonumber\\[10pt]
&+&\displaystyle{\frac{2}{\frac{E}{E_f}\omega}\int^T_0\frac{d}{dT}
\frac{\cos{\frac{E}{E_f}(\omega s)- \omega\vert\vec r(T)-\vec
r(T-s)\vert )}-\cos{\frac{E}{E_f}(\omega s+\omega \vert\vec
r(T)-\vec r(T-s)\vert )
}}{\vert\vec r(T)-\vec r(T-s)\vert }}\nonumber\\[10pt]
&-&\displaystyle{\frac{d}{dT}\frac{2}{(\frac{E}{E_f}\omega
)^2}(1-\frac{\sin{\frac{E}{E_f}\omega
(T+r(T))}-\sin{(\frac{E}{E_f}(\omega T-r(T)))}}{r(T)}}.\nonumber\\
\label{594}
\end{eqnarray}
All other formulae from the subsection A in the previous chapter are transformed in the same way:$\omega$ is rescaled
except in the measure, and the general multipier is added $E/E_f$. Note that the terms that arise from integration by
parts (boundary effects) are suppressed now even stronger-as $\omega T E/(E-\omega)$.
\par We keep the terms due to integration by parts so that we shall be able to check expicitly that they are small
in our analysis of the eq. (\ref{594}). For convienience, let us
write the latter equation without the backreaction term, that is
the result that will be used in the calculations: \beq
\displaystyle{\frac{dW}{dT}} =q^2\frac{4}{\pi}\omega d\omega
\displaystyle{(\int^T_0dt(\vec v(T)\cdot \vec v(t)-1) \times
\cos{(\frac{E}{E_f}\omega
(T-t)}}\displaystyle{\frac{\sin{(\frac{E}{E_f}\omega\vert\vec
r(T)-\vec r(t)\vert)}}{\vert\vec r(T)-\vec r(t)\vert}}. \label{s}
\eeq
\par The recoil effects lead to the qualitative change of the
spectrum of the single particle. The maximum of the radiation
reaction will bew shifted to $\omega_m\sim 0.4 E$ for large
$\chi$, and will be virtually $\chi $ independent. For the
opposite limit of small $\chi$ the maximum will remain at the
classical value of $\sim E\chi$.
\subsection{Recoil effects in the dipole radiation.}
\par It is clear from the preceding sections that taking recoil into account will mean just rescaling $\omega$ in the
previous chapter. Consequently, we obtain:
 \begin{eqnarray}
\displaystyle{\frac{dW}{d \omega dT}}
&=&\displaystyle{\frac{4q^2}{\pi}}\int^T_0dt\cos{(\displaystyle{\frac{E}{E_f}}\omega
(T-t))} (\vec v_P(T)\cdot \vec
v_P(t)-1)\displaystyle{\frac{\sin{\vert \vec r_P (T)-\vec r_P
(t))\vert}}{\vert\vec r_P(T)-\vec r_P(t)\vert )}}
\nonumber\\[10pt]
&\times&(1-J_0(\frac{E}{E_f}\omega \theta (T)d(T)))
\nonumber\\
\label{692}
\end{eqnarray}
where the function $\theta (T)$ is given by eq. (\ref{il}). Since
the main contibution still comes from $s\sim T$, we obtain:
\begin{eqnarray} \displaystyle{\frac{dW}{d \omega dT}}
&=&-\omega\displaystyle{\frac{4q^2}{\pi}}\int^T_0dt
\displaystyle{\frac{(1-v^2(T)+\omega_0^2s^2/2}{s}}
\sin{\displaystyle{\frac{E}{E_f}}(\omega
((1-v)(s))+\omega_0^2(s)^3/24))}\nonumber\\[10pt]
 &\times &(1-J_0(\displaystyle{\frac{E}{E_f}}\omega
\theta (T)d(T))).\nonumber\\  \label{683a} \end{eqnarray}
\par One can obtain the full radiation reaction by integrating the above equation over all frequencies.
\section{Radiation reaction for the very small dipole.}
\par Let us analyse the above equations for different regimes discussed in the chapter 3. We consider in this section
the case of the very small dipole:$1/\omega\ll T\ll m/F$. First,
let us check, what time scales  contribute to eq. (\ref{681}) in
this case. For the linear term in the argument of cos in eq.
(\ref{681}) to be dominant we need:
$$(1-v)\gg \omega_0^2s^2/24,$$
or
$$s\ll 2\sqrt{6}\frac{m}{\sqrt{2}E}(E/F)=2\sqrt{3}\frac{m}{F}\sim 3.5 \frac{m}{F}.$$
Here $s=T-t$, Since the latter condition is satisfied for the very
small dipole for the entire integration region in s, we can
neglect the cubic terms in the arguments of the cos as well as the
nonleading terms in the preexponentials. The integrals over $s$ in
eq. (\ref{681}) can be taken explicitly. As it is explained in the
appendix A, in this case we can discard the terms in eq.
(\ref{681} proportional to $1-v^2$, as well as the terms
originated from the integration by parts. The reason is that up to
the terms suppressed as $m^2/E^2$ these terms correspond to
radiation reaction of the free charged particle moving with a
constant velocity. The latter is of course a nonphysical phenomena
(see discussion in the Appendix A), and must be substracted. We
start with an integral
 \begin{eqnarray}
\displaystyle{\frac{dW}{d \omega}}
 &=&-\displaystyle{\frac{4q^2}{\pi}}\int^T_0dt\int^t_0ds
\displaystyle{\frac{\omega_0^2}{2}}
s\sin{\displaystyle{\frac{E}{E_f}}(\omega
((1-v)(s))}\nonumber\\[10pt]
 &\times&(1-J_0(\displaystyle{\frac{E}{E_f}}\omega
\displaystyle{\frac{m}{E}d(T)))}\nonumber\\  \label{683}
\end{eqnarray}
 Note that the latter
integral, as it is well known from the classical theory \cite{LL}
is proportional to $F^2$, i.e. to the square of the acceleration.
\par The latter integral can be taken eplicitly under the
assumption that the interference multiplier weakly depends on time
T in the limit of integration, and thus can be taken outside of
the integrand. We obtain:
\begin{eqnarray}
 \frac{dW}{d\omega}&=& \omega_0^2\omega\frac{q^2}{\pi}(
\frac{2(1-\cos{(\omega' (1+v)T})-\omega (1+v)T\sin{(\omega'
(1+v)T}
}{\omega^{'}3(1+v)^3}\nonumber\\[10pt]
&-& \frac{2(1-\cos{(\omega' (1-v)T})-\omega (1-v)T\sin{(\omega'
(1-v)T}}{\omega^{'3}(1-v)^3})((1-J_0(\displaystyle{\frac{E}{E_f}}\omega
\displaystyle{\frac{m}{E}}d(T))).
\nonumber\\
\label{699}
\end{eqnarray}
\par The corresponding spectral curve for the single particle
(without taking into account radiation) is depicted in Fig.
\ref{Fig23}. The spectral curve for the same energy and field, but
for the dipole, whose transverse motion velocity is $v_t\ll 1$ and
the interference is taken into account  is depicted in Fig.
\ref{Fig25}. We see the drastic decrease of the radiation for all
frequencies. We also see, that increasing the factor $\gamma$
(i.e. decreasing mass for given energy) leads the maximum to be
shifted further to the end-point of the spectrum (Fig.
\ref{Fig26}).
  Indeed, the
classical maximum of radiation is at \beq\omega_{\rm cl}\sim
\frac{1}{T}\frac{E^2}{m^2}\label{vs}\eeq and for sufficiently big
energies is beyond the end-point. This is the case when the recoil
effects are most important. \par Note  that at $T\sim m/F$ the
classical radiation maximum $\omega_{\rm cl}$ of eq. (\ref{vs})
reaches $\omega_H$-the classical radiation maximum for small (but
not very small) time regime.
\par We see that the effects of interference are the biggest if
the transverse motion is nonrelativistic. For $v_{0t}\sim 1$ the
interference effects are small (see Fig. [\ref{Fig27}]).
 \par The
latter equation can be
differentiated in T and then integrated over $\omega$ to obtain
the total radiation reaction of the very small dipole:
\begin{eqnarray} \displaystyle{\frac{dE}{dT}}
&=&\frac{q^2}{\pi}T^2E^2\omega^2_0\int^\infty_0dx\displaystyle{\frac{1}{x*(1+x)^3}}
\sin{(xb)}-b*x\cos{(b*x)})/b^2\nonumber\\[10pt]
&-&(\sin{(xa)}-a*x\cos{(a*x)})
/a^2)(((1-J_0(x md(T)))\nonumber\\
\label{932}\end{eqnarray} Here $a=ET(1-v),b=ET(1+v)$.
 The latter intergal, contrary to the
single particle case (see Appendix A), can't be taken explicitly.
In order to estimate this integral, it is worthwhile to get rid of
oscillating terms using the the representation
$$\frac{1}{(1+x)^3}=0.5\int^\infty_0p^2\exp{(-p)},$$
and the equations (\ref{int1}) and (\ref{int2}) from Appendix B.
We obtain \beq
\frac{dE}{dT}=0.5\frac{q^2}{\pi}\int^\infty_0p^2\exp{(-p)}
(G_1(p,b,md(T))-G_2(p,b,md(T))\label{int4}\eeq The corresponding
time dependence is given in Fig. \ref{Fig28} We put the Figure
below the radiation reaction curve for the single particle Fig.
\ref{Fig24}. The   radiation reaction decreases drastically due to
interference .
\par Finally, note that for $T\sim m/F$ radiation reaction for
$\chi\rightarrow \infty $ behaves like $1/\chi$, while for $\chi
\rightarrow 0$ like $\sqrt{\chi}$.
\par Numerically, it is easy to see that the condition for the
interference to decrease significantly the total radiation
reaction, is that the radiation maximum must occur for the
frequencies where the interference is still strong, i.e. this
frequency $\omega_m$ is such that \beq
\omega_mE/(E-\omega_m)md(T)\le 1\label{fr}\eeq It is easy to see
that this condition is equivalent to \beq E/m\le T/d(T)\sim
1/v_t\label{en}\eeq In other words, the interference decreases the
dipole radiation by order of magnitude if it is nonrelativistic in
it's c.m. reference frame, while the interference influences the
total radiation reaction only lightly if the dipole transverse
motion is relativistic ($v_{0t}\sim 1$).
\par It is worthwhile to describe qualitatively the position of
the radiation maximum and the structure of the spectral curve for
different $\chi$. First, consider $\chi\ll 1$. In this case we see
that at $T\sim E/m^2\ll m/F$ the maximum of radiation will be near
the end-point of the spectrum. The radiation itself will be
negligible. For larger times the total radiation slowly increases,
while the radiation maximum moves to $\sim E\chi$, where it
reaches at times $\sim m/F$.  Then the curve smoothly transforms
itself into the curve for small dipole that is studied in the next
section. The radiation maximum does not move anymore. Afterwards
radiation reaction quickly increases, while the interference
diminishes. The reason why for the nonrelativistic transverse
motion the radiation is still suppressed at times $\sim m/F$ is
that the suppression factor in the maximum is $1-J_0(v_{0t}x_m)$.
\par For the opposite case $\chi\gg 1$ the situation quite
different. There is a complete suppression of radiation and the
radiation maximum remains near the end-point well into the small
dipole regime (see next section).
\par The results for the radiation reaction are in correspondence with the situation with the total
number of the photons. Without interference (see Fig.\ref{Fig231})
the total number of radiated photons remains finite, and has a
maximum . Including the interference cuts off the soft photons and
decreases the total number of photons drastically (see Fig.
\ref{Fig271}). The maximum of the number of radiated photons
distribution, as is it is seen from the figure is parametrically
located at the same frequencies as that of radiation reaction,
i.e. most of radiated photons are hard.
\par In this discussion we did not take into account the important
effects of Sudakov form-factors and wave function renormalization,
that generally tend to cancel out. For the number of photons we
expect these factors be more important than for the energy.
Consequently, our discussion is just a conjecture that needs
further calculation.

\section{Small dipole.}
\par In the previous section we discussed the case of very small
dipole, corresponding to $T\ll m/F$. The goal of this section will
be to consider the opposite limiting case $T\gg m/F$. In the
latter case we can substitute the integration limits by infinity
and discard the terms due to integration by parts. We immediately
obtain for spectral density: \beq
\frac{dE}{dTd\omega}=\frac{2q^2}{\sqrt{\pi }}\frac{m^2}{E^2}\omega
( \frac{1}{2}\int^\infty_a\Phi (u)du+\frac{1}{a}\frac{\partial
\Phi (a)}{\partial a})(1-J_0(\omega' \theta(T)d(T)) \label{1377}
\eeq where \beq a=(\omega /\omega_H)^{2/3}\,\,\,
\omega_H=\omega_0(E/m)^3\label{oh}\eeq
 \beq \omega'=\omega
E/(E-\omega )\label{tt}\eeq Function $\Phi$ is the standard Airy
function (see Appendix B).
 This result is  the single particle answer times the
interference multiplier. \par Let us first consider the case of
$\chi\ll 1$. The corresponding graphs are depicted in Figs.
\ref{Fig71},\ref{Fig81},\ref{Fig91}. We put the graphs \ref{Fig81}
and \ref{Fig91} under the graph \ref{Fig71}, that corresponds to
the spectral curve of the radiation reaction for a free particle.
The graph \ref{Fig81} corresponds to time $T\sim m/F$, while
\ref{Fig91} for time $T\sim E/F$. We see that for the first graph
the interference is very strong, while for the second case it is
only slight.
\par The opposite limiting case $\chi\gg 1$ is depicted in Figures
\ref{Fig101},\ref{Fig111},\ref{Fig121},\ref{Fig131}. We see that
for $T\sim m/F$ the interference decreases dramatically the
radiation reaction, for $T\sim E/F$ the decrease is only slight
(if fact we see very slight increase). Most interesting, we see
that interference remains important numerically even at $T\sim
(E/F^2)^{1/3}$, where we see that it decreases the maximum by the
order of 1.5 and clearly significantly decreases the total
radiation reaction (the spectral curve is still "dipole-like").
Moreover, the interference leads to the further shift of the
radiation maximum to the end-point of the spectrum.
\par Let us now consider the total radiation reaction.
 Integrating eq. (\ref{1377}) we obtain:
\beq
\frac{dE}{dT}=\frac{2q^2}{\sqrt{\pi}}m^2\int^\infty_0dx\frac{x}{(1+x)^3}
(0.5\int^\infty_u\Phi(u)+\Phi'(u)/u)(1-J_0(xEd(T)\theta (T)))
\label{rr} \eeq Where the function $\theta (T)$ is given by eq.
(\ref{il}) above. The expression in the case without interference
differs slightly from the one  for the single particle, since we
did not do the usual integration by parts. \beq
u=\frac{x^{2/3}}{\chi^{2/3}}\label{fr1}\eeq
\par In order to
understand qualitatively the influence of the dipole interference
let us change the integration variable to
   u. Then \beq
\frac{dE}{dT}=\frac{3q^2\chi^2}{\sqrt{\pi}}m^2\int^\infty_0du\frac{u^2}{(1+\chi
u^{3/2})^3}((0.5\int^\infty_u\Phi(s)ds+\Phi'(u)'u)(1-J_0(u^{3/2}\chi
Ed(T)\theta (T))).\label{chi1}\eeq The radiation reaction is  the
function of two parameters: the relativistic invariant $\chi$ and
relativistic invariant $b(\tau)$ which is  equal to \beq
b(\tau)=E\theta (T)d(T)=m\frac{d d^2(\tau )}{d\tau }\label{ri}\eeq
for small dipole and \beq b(\tau )=md(\tau )\label{vs1}\eeq for
the very small dipole. The radiation reaction can be written as
\beq \frac{dE}{dT}=F(\chi, c(\tau))\label{c}\eeq where \beq
c(\tau)=b(\tau)\chi .\label{ist}\eeq Here $\tau$ is the proper
time in the c.m. reference frame of the dipole.

\par In order to
understand qualitatively the influence of the dipole interference
consider two limits: $\chi\ll 1$ and $\chi\gg 1$.
 For $\chi\ll 1$ we can use the equation (\ref{chi1}). For the time
 $T\gg m/F$ we can estimate:
 $$d d^2(T)/dT\sim F^2T^3/E,$$
 and the argument of the Bessel function in eq. (\ref{chi1}) is just
 $u^{3/2} (T/T_0)^3$, where $T_F=m/F$. We then put $\chi =0$ in
 the denominator in the latter equation and obtain:
 Thus we obtain
 \beq
 dE/dT=-\frac{2}{3}\frac{F^2E^2}{m^4}G(T/T_F),\label{Pom}\eeq
where the first term corresponds to the classical Pomeranchuk
effect, and the function $G(T/T_F)$ is defined by \beq G(s)=
\frac{9}{2\sqrt{pi}}\int^\infty_0
u^2(0.5\int^\infty_u(\Phi(u)+\Phi' (u)/u)(1-J_0(u^{3/2}(T/T_F)^3)
\label{G}\eeq It is clear that for $T\gg T_0$ $G(T)\rightarrow 1$.
The radiation reaction in this case is depicted in
Fig.\ref{Fig202}, where we see, the sharp increase of radiation
reaction at $T\sim T_F$. We see that it becomes weakly time
dependent numerically at $T\sim E/F$.
\par For the opposite case $\chi\gg 1$ the main contribution in
the integral of eq. (\ref{chi1}) comes from the $u\rightarrow 0$.
In this case we can put the argument in the Airy functions to
zero, and use
 $\Phi'(0)\sim -0.5$. We then obtain
 \beq
 dE/dT=-\frac{q^2m^2\chi^{2/3}\Phi'(0)}{\pi}F(T/T^*),\label{chi2}\eeq
 where
\beq F(T/T^*)=\int^\infty_0x^{1/3}/(1+x)^3
(1-J_0(x(T/T^*)^3)).\label{chi3}\eeq i.e. Here
$T^*=(E/F^2)^{1/3}$.
\par Note that for $\chi\gg 1$ $E/F\gg T^*\gg T_F=m/F$. Thus the
argument of the  Bessel function x is multiplied by  a number less
than 1.
 Consequently, as it is
clear from the Figure \ref{Fig203}, the radiation reaction remains
negligible up to $T\sim T^*$. Afterwards, it increases rapidly,
and at $T\sim E/F$ becomes approximately time-independent and a
sum of radiation reactions of the components of the dipole.

\par In both cases the qualitative dependence on the parameter
$\chi$ for the dipole is the same as for the single particle.
\par It is worthwhile, as in the previous section, to follow the
position of the maximum of radiation. For small $\chi$ as we saw
it does not change, and the interference in this regime is small.
The interesting case is the case of the large $\chi$. We see that
up to $T\sim T^*$ the spectrum is shifted to the end-point. This
is in contrast with the single particle, where, as it seen from
Fig. \ref{Fig101}, (see also refs. \cite{AS1,AS2}) the spectrum is
concentrated near $\omega_m\sim 0.4E$. The strong interference in
the maximum is the reason of the suppression of the total
back-reaction. When $T\ge T^*$ the maximum, as it is clear from
the Figures above, starts to move to the position of the maximum
of the single particle, i.e. 0.4E, where it comes by $T\sim E/F$,
and the radiation reaction quickly increases.
\par Consider now  the number of radiated photons
in $q^2$ approximation of the perturbation theory. It is clear
that this number  is drastically decreased by interference. We see
in Fig. \ref{Fig11} the number of emitted photons without the
interference, while in Fig. \ref{Fig12} the number of emitted
photons is shown with the  interference taken into account. We see
that the interference qualitatively changes the spectrum: instead
of being infinite in the limit of soft photons ($\omega\rightarrow
0$), now the spectrum has no infrared singularity for soft
frequencies. Instead a number of radiated photons $\rightarrow 0$
at $\omega\rightarrow 0$ and has a finite maximum at finite
frequency. Moreover, it is clear that up to numerical coefficient
of the order one, the position of this maximum will be the same,
as the  frequency that corresponds to the maximum of the radiation
reaction. In particular, for ultra-relativistic dipole $\chi\gg 1$
the relevant frequency will shift to the endpoint of the spectrum.
The photons will take almost the entire energy of the radiating
electron in a single radiation event for the time interval $T^*\ll
T\ll E/F$. For $\chi\sim 1$ one radiating event will take $\sim
1/2$ of the initial energy of the electron.
\par Even more interesting effect will take place for higher times
$T\sim E/F$ (and $\chi\gg 1$). Since the soft photons will be cut
by the dipole effects, the number of photons will have the maximum
 for the finite  frequencies.
Numerical analysis shows that in this case \beq \omega_m\sim
2/T\label{nmb}\eeq This means that there are two distinct groups
of radiating events for  large T: radiation of large numbers of
soft photons with frequencies given by eq. (\ref{nmb}) and the
radiating events where the dipole loses approximately half of it's
energy each time, this half being carried by a photon.
\section{Back-reaction and evolution of the very small dipole}
\par We can answer now how the back-reaction influences the evolution
of the very small dipole, and where does the energy loss due to
radiation goes: to the relative motion of the particles in the
center of mass or to the loss of the total energy of the center of
mass motion.
\par Our results show, that for very single particle and for the very  small
times the back-reaction force behaves according to eq.
(\ref{force}) in the Appendix A. For the dipole the back-reaction
force behaves much less singular. Note that for very small times
one can write \beq dE/dT=\frac{q^2a^3}{\pi (1-v)^2}\int^\infty_0ds
(\sin (s)-s\cos (s)*(1-J_0(2v_{0t}s)/(s(s+a)^3).\label{exp}\eeq
Here as in the previous sections $a=ET(1-v)$. The equation
(\ref{exp}) can be used to obtain the first several terms in the
expansion of the back-reaction force for small T: \beq dE/dT=
\frac{q^2T^3m^2F^2v_{0t}^2}{2\pi
E}(1-\frac{3}{2}\frac{m^2T\pi}{4E})+0(T^5\log{(m^2T/E)}\label{det}\eeq
Recall that $v_{0t}$ is the transverse velocity in the c.m. frame.
\par The back-reaction force for the dipole is smaller than for the
single particle, logarithmic terms are present only starting from
$T^5$, and it's leading term is proportional to $T^3$. The
condition for the applicability of the expansion (\ref{det}) is
 \beq T\ll {\rm min} E/m^2, m/F .\label{gvul} \eeq
\par Since the photons are radiated almost parallel to the direction of the
dipole, the corresponding back-reaction force is directed in the
direction opposite to the direction of the dipole and leads to the
decrease of it's center of mass velocity.
\par There is an additional back-reaction force, that slows the expansion
of the dipole in the orthogonal direction. For very small times
regime this force is evidently $\sim \sin^2{\theta}dE/dT/p_t$,
where $\theta\sim m/E$ is the radiation cone angle, and $p_t\sim
v_{0t}m$ is the transverse momentum. Consequently the orthogonal
component of the back-reaction force is \beq F_y(T)\sim
\frac{q^2}{\pi}\frac{F^2m^3T^3v_{0t}}{E^2}+0(T^4,T^5\log{(m^2T/E)}.
\label{back} \eeq
 The influence of this force on the wave packet
radius begins only from the terms of $\sim T^5$, i.e. for very
small T the expansion due to quantum diffusion and external field
is dominant.

\section{The Quantum dipole.}
\par In the latter analysis we did not take into account the quantum character of the dipole motion. In fact, there
is an additional effect that influences the motion of dipole, and
this is the noncoulombic photon exchange between the components of
dipole. This effect is significant if the distance between the
components of the dipole is less then $1/m$, and leads to so
called quantum diffusion: the distance between the components of
the dipole increases not linearly or quadratically as in the usual
relativistic quantum mechanics, but in the diffusion way, i.e. as
$\sim \sqrt{T}$ \cite{Farrar}
( a simple qualitative explanation of this phenomena is contained
in ref. \cite{Doc}). 
Moreover, the dipole motion 
along the coherence length may stop to be quasiclassical, as it was
assumed throughout this paper \cite{Farrar}.
\par It is easy to see that the effect is important for
ultra-relativistic dipole with $\chi\gg 1.$ Indeed, the diffusion
is important till the distance between the components of dipole is
$\sim 1/m$, where $1/m$ is the scale of bound state in QED. In
order to take into account the external field we need to write the
wave functions in the external field taking into account quantum
noncoulombic exchange of Fig. \ref{Fig16}. This is beyond the
scope of the current paper. Here we shall try to build a
qualitative model to indicate the influence of the quantum
effects. In order to estimate the field influence on the diffusion
let us note that the diffusion law, \beq
d^2(T)=\frac{2T}{E},\label{DL}\eeq can be obtained from the
equation \beq \dot{y}/E=\eta (t),\label{rf}\eeq where $\eta$ is
the random external force such that: \beq <\eta (t)\eta
(t')>=E\delta (t-t').\label{cor}\eeq In order to include the
external field, we generalize this equation in an obvious way:
\beq \ddot{y}(t)+\dot{y}/E=\eta (T)+F/E\label{exteq}\eeq This
equation can be easily solved with the result: \beq
d^2(T)=4<y^2(T)>=2T/E+2F^2T^2/E^4+O(T^3),\label{d22}\eeq and \beq
<y(T)>=FT/E^2\,\,\,\,<v_y(T)>=F/E^2\label{dop}\eeq We see that
quantum diffusion changes the velocity and the distance between
charges. Average velocity is small $F/E^2$ and constant. Diffusion
is important till the distance between dipole components is $1/m$,
i.e. $2T/E\sim 1/m^2$, or \beq T\sim (E/m)(2/m)\label{T1}\eeq Note
that for $\chi\gg 1$ this time is $\gg m/F$. Also note that for
all reasonable times $\le E/F$ the first term in eq. (\ref{d22})
is dominant.
\par Consider now the interference in the case of diffusion.
We need to average the product $\sin{(\theta )}\omega d$. There
are two possibilities. If the angle between the direction of a
component of the dipole and z axis is much bigger than $m/E$, we
use $\sin\theta\sim v_y$. Then we need to average $<v_yd(T)>=
\displaystyle{\frac{d d(T)^2}{dT}}=2/E$, and we get as the
argument of the Bessel function $2x=2\omega /E$ ($\omega'/E$ if we
also take into account recoil effects) . The interference
multiplier will be \beq (1-J_0(2x)).\label{IM}\eeq If the angle is
$m/E$ we shall get the argument $\omega m<d(T)>/E\sim 2\omega
mFT/E^3$. We must choose the biggest of two arguments. It is easy
to see that the first argument will be bigger up to times $T\sim
E^2/(mF)=(E/F)(E/m)$, i.e. for all times where the dipole notion
has sense. Thus, in our simple model, in the diffusion regime,
which lasts parametrically longer, as $\chi$ increases, the
interference depends on time only weakly. \par  For small $\chi$
the diffusion law holds only for very small times $\ll m/F$. Thus
the biggest influence seems to occur for $\chi\ge 1 $, when the
interference multiplier may  significantly change the radiation.
\par We have developed above a simple phenomenological model
indicating  are the effects connected with the quantum character
of the dipole motion. Unfortunately, at the moment we can, on the
basis of this model, only indicate, that they may be very
important for  large $\chi$ and that they lead to the suppression
of the dipole radiation , as in the quasi-classical dipole.
\par The reason of the difficulties we encounter is the inadequacy
of the classical approximation. It is possible to estimate the
area of reliability of the quasi-classical approximation: we must
demand that the transverse velocity acquired in the classical
approximation due to the action of the external field is bigger
than the velocity due to the quantum diffusion. Quite, remarkably,
this gives us the condition $T\ge T^*$, i.e. the quantum diffusion
effects are important for large $\chi$ up to the scale, when
quasi-classically radiation suppression stops, and the radiated
energy quickly increases as we saw in the Chapter 4. This result
is consistent with the conclusion above, that in the time interval
when the quantum effects in the dipole motion are important, there
is still a suppression of the radiation reaction (the charge
transparency). The origin of the difficulty is clear. The
parameter $\chi=l_c/l_F$, where $l_c\sim E/m^2$ is the coherence
length, while $l_F\sim m/F$ is the field regeneration length, in
other words the average distance the dipole must travel before
colliding with the external field photon. It is clear that the
external field does not break coherence. Thus we are in the
situation when we have multiple coherence conserving collisions
along the coherence length. In this situation it is well known
(see e.g. ref. \cite{Imri} ) that the classical approximation is,
generally speaking, not applicable. The quasiclassical
approximation corresponds to neglecting the coherence conservation
and thus can lead to the wrong results. The further analysis along
the lines of ref. \cite{Imri} is needed.
 \section{Conclusion}
\par We have studied the back-reaction and it's influence on
the evolution of the relativistic dipole in the arbitrary strong
external field using the quasi-classical approximation. We have
taken into account the quantum recoil effects in radiation, but
not quantum effects in the motion of the dipole,i.e. the quantum
diffusion. We found that the dipole motion is governed by two
invariant parameters, one of them describes the longitudinal
motion and is equal to $\chi =EF/m^3$, another describes the
motion in the transverse plane and is equal to
$$b(\tau )=md(\tau)\,\,\,{\rm if }T\ll m/F$$
$$b(\tau )=m\frac{d d^2(\tau)}{d\tau}\,\,\, E/F \gg T\gg m/F.$$
It is quite possible that there exists a single formula for b,
although we were not able to obtain it.
\par We have studied the pattern of charge transparency in the external field.
We have  found that the interference
 effects can be taken into account by the use
of the  general interference multiplier $1-J_0(xb(\tau)$, where
$x=\omega/(E-\omega )$. For the arbitrary times the radiation
reaction is given by eq. (\ref{681}).
\par We have seen that there
are three different time scales. First, the very small dipole
regime, $T\ll m/F$. This time scale exists if the dipole
transverse velocity $v_{0t}\ll 1$. In this regime the radiation
reaction is strongly suppressed by interference, leading to the
strong decrease of the back-reaction, i.e. the fast moving dipole
does not loose it's energy. In this case we were able to calculate
analytically the back-reaction force analytically for both the
entire regime, and for very small times (eq. (\ref{det})). For
larger time scales $E/F\gg T\gg m/F$ the influence of interference
on back-reaction depends on the value of the parameter
$\chi=EF/m^3$. If $\chi\ll 1$, the radiation reaction quickly
increases starting from $T\sim m/F$, and by the time $T\sim E/F$
it is a sum of radiation reaction of the components of the dipole
(see Fig. \ref{Fig202}). However  for the opposite case $\chi\gg
1$, the radiation reaction starts to increase only from the times
$T\sim T^*=(E/F^2)^{1/3}$, and then once again goes quickly to the
sum of the radiation reactions of the components.
\par For the third regime $T\gg E/F$ the components of the dipole
can be considered as independent particles. For each of the
regimes we obtained the analytical expressions for both the
spectral distribution of the radiation and the total back-force.
The results are qualitatively shown in Figures 1-20.
\par The results for the radiation reaction are in the
correspondence with the influence of the interference on the
number of radiated photons and on the scattering cross-sections.
These physical quantities are qualitatively influenced by
interference up to times $\sim E/F$. Without the interference the
number of photons is maximum (infinite) at $\omega\rightarrow 0$.
As a result of the interference the number of photons goes to zero
when $\omega\rightarrow 0$. The number of photons distribution has
maximum at $\omega^*\sim \gamma /d(T)$ for the very small dipole
regime. I.e. for ultra-relativistic dipole the maximum of the
number of radiated photons shifts to the end-point of the
spectrum. \par For larger times (small dipole regime) the maximum
in the number of the radiated photons is finite and parametrically
lies at the same frequencies as the maximum of the radiation
reaction -$E\chi$ for $\chi\ll 1$, 0.4E for $\chi\gg 1$. Moreover,
we have seen that for $\chi\gg 1$ the radiated particles carry
$\sim 1/2$ of the dipole energy for arbitrary times $T\gg E/F$,
when the particles move as the independent ones. There are two distinct
maximums and two groups of photons. One group is responsible for the
energy loss, and it's spectral curve maximum is at $\omega\sim 0.4E$ for
large $\chi$. Another group is the soft photons, responsible for a
total number of photons emitted (and they may give the main
contribution into the cross-sections). These photons in the regime
under discussion are the soft ones, with the maximum of the
spectral curve located  at $\omega\sim 2/T$ for large $\chi$.
 \par It is also interesting to summarize the behaviour
of the radiation spectrum for different $\chi$. For $\chi\ll 1$, the
relevant maximum lies near end-point if $T\le E/m^2$, but the
radiation is strongly suppressed. However if it occurs, the dipole
will be immediately destroyed, since photon takes all of it's
energy. Then it moves to $E\chi$ by the time $m/F$, and only
afterwards the radiation begins to increase.
\par In the opposite case $\chi\gg 1$ the maximum is near
end-point till $T\sim T^*$, and only then begins to move to
saturation 0.4E, that corresponds to the single particle maximum.
For $T\le T^*$ the radiation is suppressed, but if occurs it
destroys the dipole (photon carries it's entire energy). For the
times $T\gg E/F$ the radiation reaction is a sum of the component
radiation events, and at each radiation event on average half of
the electron energy is taken by the photon.
\par We have seen, that our results, although they were obtained
for the simple model of the constant transverse field, can be
reformulated in the model independent way. The parameter
$\chi=l_c/l_F$ is the ratio of the coherence and the field
generation length. The very small dipole regime corresponds to the
situation when the dipole travels the distance less than $l_F$.
There exists the charge transparency in this region independently
of $\chi$. However for large times the parameter $\chi$ starts to
play an important role. If $\chi\ll 1$, (this is the situation
considered in refs. \cite{Farrar,FS1,FS}) one can see the external
field as a small perturbation. The region of the quantum diffusion
is small (since $l_c\ll l_F$) and charge transparency (i.e.
radiation suppression) continue up to $T\sim m/F$), well beyond
the quantum diffusion range. However, once $\chi\ge 1$, we are in
a completely different situation. In the quasi-classical approach
we have here the situation quite similar to the Landau-Pomeranchuk
effect in the single particle dynamics of the fast particle moving
through the amorphous media. The radiation reaction continues to
be suppressed even after the field regeneration time, thus
extending parametrically the charge transparency interval to times
$T\sim T^*$ . However, as it was noted in the previous section,
the area of $T\le T^*$ must be studied beyond the quasi-classical
approximation, since we must take into account the multiple
coherent collisions. We expect that the radiation will still be
strongly suppressed in this time interval, but further analysis is
needed to make the qualitative statements, and to compare the
results with those from the quasi-classical approach.
\par The main possible drawback of our paper is the validity of the 
quasiclassical approximation. For the $\chi\le 1$ the quasiclassical 
approximation works for all times larger than $T_c\sim E/m^2\le T_F$.
For $\chi\ge 1$ for dipole one may expect the significant corrections 
to quasiclassical approximation for all times in light of the results of 
ref. 
\cite{Farrar} (see also the previous section) . Nevertheless the 
quasiclassical analysis is still important as a first step to 
understand the radiation patterns in this regime of parameters.
\par Our work certainly makes a number of questions open. First,
this is the influence of the quantum effects in the dipole motion
on the radiation reaction. This is important for the study of the
quantum dipole. We have seen that such effects for the dipole may
be much more important than for a single particle, and may require
the analysis beyond the quasi-classical approximation due to the
coherent multiple scattering.
\par Second, it will be interesting to study further the dependence
of the number of the radiated photons on $\chi$, in particular
taking into account the multiple photon radiation. Our results
suggest that, since the electrons are born in pairs, i.e. as
dipole, the infrared photons are always cut off, and the evolution
goes on by a series of radiative events, such that in each of
these events the electron loses approximately half of it's energy.
This is true at least if $\chi\gg 1$, i.e. the dipole is
ultra-relativistic or the field is very strong. This is opposite
to the scenario when the fast electron loses it's energy by
radiating the soft photons, with small energies loss in each of
the radiating events. This result can be important for carrying
the next-to-leading order logarithmic calculations. The results of
this paper imply, roughly, that such dipole moves till times $T^*$
without radiation, then after the transition period (up to $T\sim
E/F$), starts to radiate loosing at each event $\sim 1/2$ of it's
energy.
\par Moreover, the numerical analysis shows, that for large times there
are two parallel process for ultra-relativistic dipole. First, it
emits soft photons. The maximum of the photon number distribution
for large $\chi$, as numerical analysis of eq. (\ref{683}) shows,
lies at \beq \omega\sim 2/T, \,\,\,\,T\gg E/F.\label{na}\eeq These
photon numbers make significant contribution to cross-sections.
But the energy loss of the dipole goes via the series of different
events, when $\sim 0.4E$ is lost in each event, and the relevant
photons are hard for the ultra-relativistic dipole.\par It will be
interesting to understand if the different regimes of radiation
discussed in this paper are connected with the theory of the
production of the $e^+-e^-$ pairs by fast particle in the external
field discussed in ref. \cite{M1}.
\par Finally, it will be interesting to study the implications of
our results for QCD. In particular, our results are clearly
relevant to the studies of the  colour transparency phenomena,
first discussed in refs. \cite{Farrar,FS1,FS}. As it was noted in
the introduction for the case of the deep -inelastic scattering on
the longitudinal photons the charge transparency is directly
translated into the color transparency \cite{CFS}. Our results
give qualitative bounds on the color transparency for the
arbitrary external fields, and indicate the direction of the
research one needs to extend the color transparency to the case of
the arbitrary external field.
\par It will be especially interesting to extend our results to
the gluon color dipole radiation, since then the shift of the
spectrum to the end-point will mean that the dipole loses all it's
energy by a single radiative event for small time. It will also
imply that the color dipole loses it's energy by a series of
events in each of them the gluon looses half of it's energy. Note
however that the extension to color dipole is nontrivial since the
mass of the gluon is zero, and we need the additional
regularization. Moreover,  the definition of QCD dipole is 
slightly different then the one in this paper. In this paper 
dipole is a system of two oppositely charged particles with 
interfernce, QCD dipole is only a quantum dipole, i.e. the 
times considered are always less than the time interval that 
corresponds to the coherence length. For such case, as we saw 
above there may be significant corrections to the 
quasiclassical approximation that need further study. Nevertheless our 
results imply that the recoil
effects may be very important also for the color dipole, i.e. for
the small x deep -inelastic scattering.
\newpage
\acknowledgements{The author thanks Professor L. Frankfurt for
numerous useful discussion and reading the manuscript}
 \appendix
 \section{Radiation reaction for single quantum particle for small times
 (small deflection angles).}
\par Although the article is devoted to the radiation reaction of a
dipole, in this section we shall discuss the radiation reaction of
a single particle for the small times, taking into account recoil
effects. Although such problem may look un-physical, since charged
particles are created by pairs, exactly the same problem appears
if the particle goes through the line of the constant external
field of the length $L\le m/F$. The purely classical case was
discussed by Landau and Lifshits \cite{LL}.However we were not
able to find the quantum case in the literature.
\par We start from eq. (\ref{501}) for total energy radiated by a
single particle during the time interval T. Using the
approximation of eqs.  (\ref{final}) and (\ref{681})  we see that
the radiation reaction is the sum of three terms: the term
proportional to $1-v^2$, the term proportional $\omega_0^2$, and
the term due to the integration by parts. Note that all cubic
terms in the arguments are negligible and thus can be omitted.
 The terms that arise due to the integration by
parts do not depend on the external field, up to the terms
additionally suppressed as $m^2/E^2$, and are the same as for the
free particle. So only the term proportional to $\omega_0 ^2$
remains. It is straightforward to find: \beq \frac{dW}{d\omega}=
\omega_0^2\frac{q^2}{2\pi}\int^T_0\int^T_0
dsds'\int^1_{-1}d\cos{\theta}) (s-s')^2\cos{(\omega
(s-s'))}\exp{i( \omega v (s-s')\cos{\theta}) )}.\label{k}\eeq The
latter triple integral can be easily taken. We obtain
\begin{eqnarray}
 \frac{dW}{d\omega}&=& -\omega_0^2\frac{q^2}{v\pi}
\frac{2(1-\cos{(\omega (1+v)T})-\omega (1+v)T\sin{(\omega (1+v)T}
}{\omega^2(1+v)^3)}\nonumber\\[10pt]
&+&\omega_0^2\frac{q^2}{\pi} \frac{2(1-\cos{(\omega
(1-v)T})-\omega (1-v)T\sin{(\omega (1-v)T} }{\omega^2(1-v)^3)}.
\nonumber\\
\label{690}
\end{eqnarray}
This is the classical formulae. the recoil is taken by first
rewriting: $$ \frac{d\omega}{\omega^2}=\frac{\omega
d\omega}{\omega^3}.$$ Then we need to rescale $\omega$ in the
r.h.s. of eq. (\ref{690}), except in the product $\omega d\omega$,
as discussed in the text: \beq \omega\rightarrow \omega'=\omega
E/(E-\omega )\label{resc}\eeq We obtain:
\begin{eqnarray}
 \frac{dW}{d\omega}&=& -\omega_0^2\frac{q^2}{v\pi}\omega
\frac{2(1-\cos{(\omega' (1+v)T})-\omega' (1+v)T\sin{(\omega'
(1+v)T}
}{\omega^{'}3(1+v)^3)}\nonumber\\[10pt]
&+&\omega_0^2\frac{q^2}{\pi}\omega \frac{2(1-\cos{(\omega'
(1-v)T})-\omega' (1-v)T\sin{(\omega' (1-v)T}
}{\omega^{'3}(1-v)^3)}.
\nonumber\\
\label{790}
\end{eqnarray}
The typical  spectral curve is depicted in Fig \ref{Fig23}.  (In
the figure we added also the contribution of the term proportional
to $(1-v^2)$). \par It is straightforward  to integrate the latter
equation over $\omega$ from 0 to E (for this we change the
integration variable to $y=\omega/(E-\omega)$. After trivial
integration we obtain:
\begin{eqnarray}
\frac{dE}{dT}&=&\frac{q^2}{\pi
(1-v)^2}\omega_0^2(((1-v)^2/(1+v)^2)(-b+{\rm Ci}(b)(b\cos b-\sin
(b)+b^2\sin{(b)}/2-b^3\cos{(b)}/2)\nonumber\\[10pt]&+&{\rm Si}(b)(\cos
(b)-b\sin{b}-b^2\cos{(b)}/2-b^3\sin{(b)}/2)+ \nonumber\\[10pt]
&+&\pi
b(-1/2+b^2/4)\sin{b}+\pi(-1/2+b^2/4)\cos{b}\nonumber\\[10pt]
&-&(-a+{\rm Ci}(a)(a\cos a-\sin
(a)+a^2\sin{(a)}/2-a^3\cos{(a)}/2)\nonumber\\[10pt]&+&{\rm Si}(a)(\cos
(a)-a\sin{a}-a^2\cos{(a)}/2-a^3\sin{(a)}/2)+\nonumber\\[10pt]
 &+&
\pi
a(-1/2+a^2/4)\sin{a}+\pi(-1/2+a^2/4)\cos{a})).\nonumber\\
\label{de}
\end{eqnarray}
Here \beq a=(1-v)ET\sim m^2T/(2E)\,\,\, b=(1+v)ET\sim
2ET.\label{ab}\eeq The corresponding typical radiation reaction
curve is depicted in Fig. \ref{Fig24}. \par  Expanding the latter
equation in powers in T we see that the back-reaction force for
small times is very small: \beq \frac{dE}{dT}=
-\frac{q^2}{6\pi}\frac{m^2F^2T^3}{E}(\log{(m^2T/2E)}+\gamma +1 )
+O(T^4), \label{force}\eeq where $\gamma\sim 0.55$ is the Euler
constant. This force is directed against the direction of the
particle.
\par The latter equation works for the whole range of
$T\le m/F$ if $\chi \gg 1$. For $T\sim m/F$, $\chi\ll 1$, the
parameter $a\sim 1/\chi\gg 1$ and we have to use the whole
equation (\ref{de}) in the limit $a\gg 1$. For $\chi\gg 1$ the
expansion parameter $a\sim m^2T/E\sim 1/\chi$ and is still small
for $T\sim m/F$. Then the back-reaction force is for this time
scale:
$$
F_{\rm b.r.}\sim \frac{q^2m^5}{6\pi EF
}\log{(E^2/m^2)}=\frac{q^2m^2}{6\pi\chi}\log{1/\chi}.
$$
 We see that the back-reaction is strongly suppressed for single
 particle in the ultra-relativistic case.
 \par Finally, let us make a comment on the discarded terms
 proportional to $(1-v^2)$, and those originated from integration
 by parts. It may be strange from the first sight that these terms
 really exist, since they are nonzero for a particle moving with
 constant velocity and finite mass, i.e. a particle that does not
 emit any radiation field. In fact this is situation usual in
 quantum mechanics and quantum field theory. Indeed, when we
 calculate the transition rate due to photon radiation in standard
 perturbation theory between stationary states we encounter
 multiplier
 \beq \sin^2(E_f-E_i-\omega)T/T(E_f-E_i-\omega)\label{T}\eeq
For infinite T this term gives a delta function $\delta
(E_f-E_i-\omega)$, insuring the law of the energy conservation.
However for finite T and non-finite $\Delta E=E_f-E_i-\omega$ this
will be a function of T decreasing as a function of T for fixed
$\Delta E$. This decrease, as it is well known, just expresses the
energy uncertainty principle. If we consider the system for finite
time, the energy can't be measured unambiguously:$\Delta E T\ge
\hbar$. The discarded terms in the radiation reaction have exactly
the same origin and the same character. They decrease for large T
as $1/T$ or faster and thus disappear at infinite T altogether.
They exhibit the ambiguity in the measurement of the
electromagnetic field of the free particle due to the finite time
of our process. In practice this leads to the finite width of
spectral lines for finite time. It is interesting to study these
terms in more detail in connection with the Landau-Pierels
inequalities \cite{LP}. However, it is clear from above, that
these terms must be discarded if we are interested in the
radiation in external field. In other words, all quantum
calculations must contain renormalization, meaning that a free
particle does not radiate.
\section{some useful integrals and their properties}
\par Here we shall collect together some useful integrals and asymptotic
expansions, as given in refs. \cite{AS,BE,WW,W}. We shall also
collect the definitions of several special functions that differ
in the literature by normalization constants. We use the following
integrals, directly expressible through Airy functions:
 \par \beq
\frac{1}{\sqrt{\pi}}\int^\infty_0ds
s\sin{(as+s^3)}=-\frac{d}{da}{\rm Ai}(a). \label{C1} \eeq \beq
\int^\infty_0ds
\frac{1}{\sqrt{\pi}}\sin{(as+s^3)}/s=-\int^\infty_adz{\rm Ai}(z).
\label{C2} \eeq Here $\rm Ai$(z) is an Airy function \cite{AS,BE}:
\beq \int^\infty_0 ds \frac{1}{\sqrt{\pi}}\cos{(as+s^3/3)}={\rm Ai
(a)}. \label{C3} \eeq  Note that  Airy function decreases as $\sim
\exp{(-z^{3/2})}/z^{1/4}$ for the positive $z\rightarrow \infty$.
\par We use integral of the Airy function: \beq
\int^\infty_0z^{b-1}{\rm Ai}(z)dz=3^{(4b-1)/6-1}\Gamma (a/3)\Gamma
((a+1)/3). \label{C4}\eeq
\par We define the Integral sinus and cosinus as:
\beq {\rm Si}(x)=\int^x_0\frac{\sin (x)}{x},\label{si}\eeq
 \beq{\rm
Ci}(x)=-\int^\infty_x\frac{\cos (x)}{x},\label{ci}\eeq
\par While studying the dipole radiation we used some formulae for the integrals
 of Bessel functions \cite{DP}.We use
\beq G_2(p,a,b)=\int^\infty_0\exp{(-px)}\sin{(bx)}J_0(ax)/x={\rm
arcsin}(2b/r), \label{int1} \eeq \begin{eqnarray} G_1(p,a,b)&=&
\int^\infty_0\exp{(-px)}\cos{(bx)}J_0(ax)\nonumber\\[10pt]
&=&
\displaystyle{\frac{1}{\sqrt{p^2+(b+a)^2}*\sqrt{p^2+(b-a)^2}}}*\sqrt{(r^2/4-b^2)}.\nonumber\\
 \label{int2} \end{eqnarray}
 Here
 \beq
 r=\sqrt{(b+a)^2+p^2}+\sqrt{p^2+(b-a)^2}.
 \label{int3}
 \eeq

\begin{figure}[htbp]
\centerline{\epsfig{figure=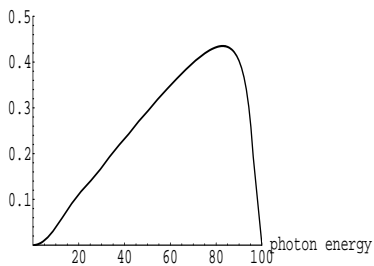,width=5cm,height=5cm,clip=}}
\caption{The  spectral curve for the spectral distribution of the
radiated energy of the single particle $\frac{dW}{d\omega}$
(normalised by $q^2/\pi$) versus the photon energy $\omega$ for a
single particle. The velocity v=0.99, $E=100$ GeV, $m=14$ GeV, The
field $F=100$ GeV$^2$, $T=0.1$ GeV$^{-1}$. } \label{Fig23}
\end{figure}
\begin{figure}
\centerline{\epsfig{figure=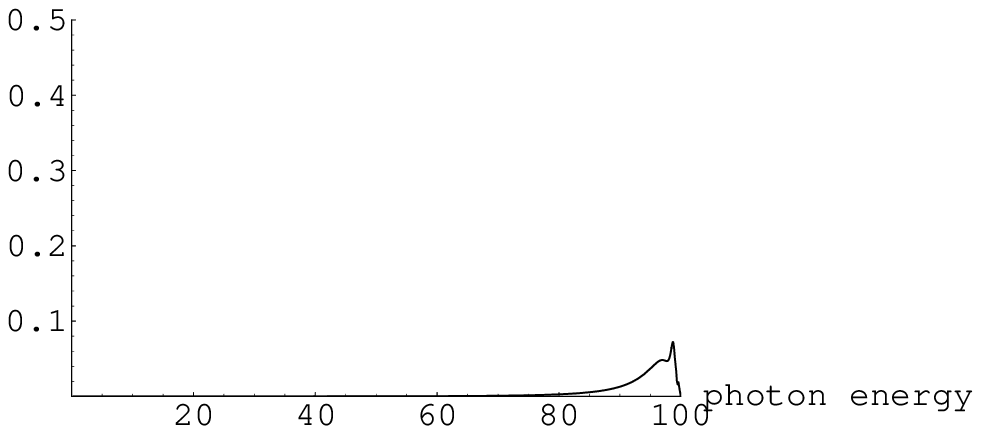,width=5cm,height=5cm,clip=}}
\caption{Spectral curve as a function of $\omega$ for the very
small time regime $T\le m/F$. The parameteres are the same as in
the previous picture:E=100 GeV, F=100 GeV$^2$,v=0.99. The
transverse velocity in the c.m. frame of the dipole is
$v_{0t}=0.2$ .} \label{Fig25}
\end{figure}
     \begin{figure}[htbp]
\centerline{\epsfig{figure=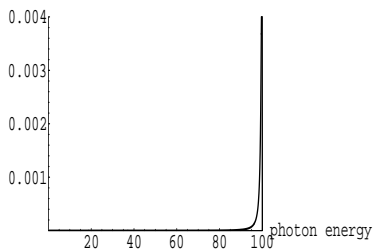,width=5cm,height=5cm,clip=}}
\caption{Spectral curve as a function of $\omega$ for the very
small time regime $T\le m/F$. The parameteres are: E=100 GeV,
F=100 GeV$^2$,v=0.999 ($\gamma =0.04$). The transverse velocity in
the c.m. frame of the dipole is $v_{0t}=0.2$. } \label{Fig26}
\end{figure}
    \begin{figure}[htbp]
\centerline{\epsfig{figure=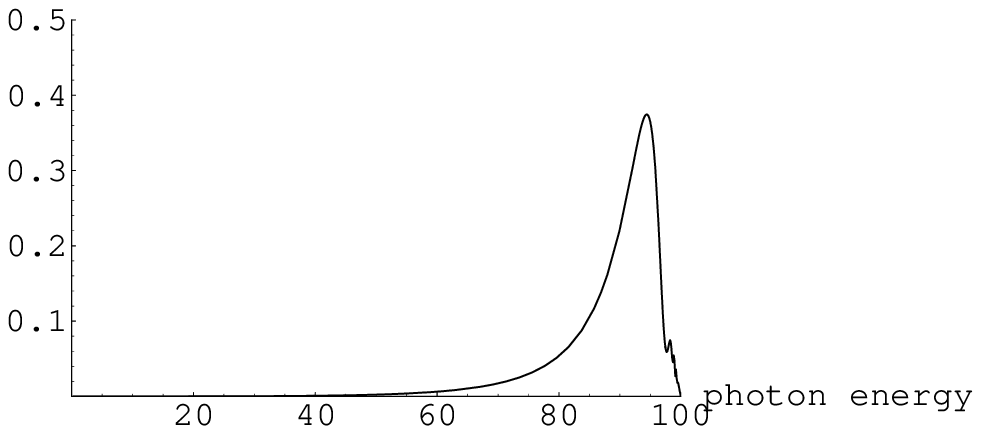,width=5cm,height=5cm,clip=}}
\caption{Spectral curve as a function of $\omega$ for the very
small time regime $T\le m/F$. The parameteres are the same as in
the Fig. [\ref{Fig25}]:E=100 GeV, F=100 GeV$^2$,v=0.99. The
transverse velocity in the c.m. frame of the dipole is
$v_{0t}=0.9$. } \label{Fig27}
\end{figure}
    \begin{figure}[htbp]
\centerline{\epsfig{figure=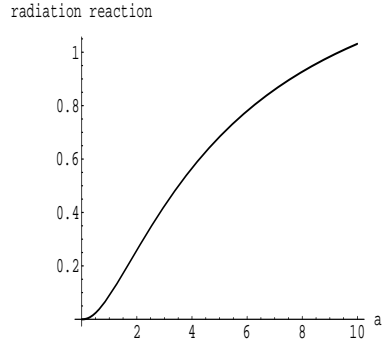,width=5cm,height=5cm,clip=}}
\caption{Radiation Reaction of the single particle as a function a
of $a=m^2T/2E$ for very small time regime $T\le m/F$. For the
picture $m^2/2E=1$ GeV, E=100 GeV, F=100 GeV$^2$. The radiation
reaction $dE/dT$ is normalized by $q^2/\pi$.} \label{Fig24}
\end{figure}
     \begin{figure}[htbp]
\centerline{\epsfig{figure=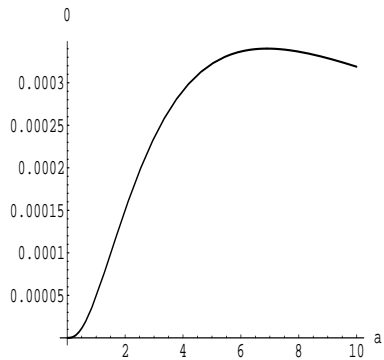,width=5cm,height=5cm,clip=}}
\caption{ Radiation Reaction of the dipole as a function of
$a=m^2T/2E$ for very small time regime $T\le m/F$. For the picture
$m^2/2E=1$ GeV, E=100 GeV, F=100 GeV$^2$. The radiation reaction
$dE/dT$ is normalized by $q^2/\pi$, and $v_{0t}=0.2$. }
\label{Fig28}
\end{figure}
\begin{figure}[htbp]
\centerline{\epsfig{figure=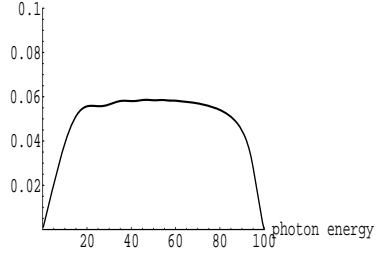,width=5cm,height=5cm,clip=}}
\caption{The  spectral curve for the spectral distribution of the
number of the radiated photons for the single particle
$\frac{dN}{d\omega}$ (normalized by $q^2/\pi$) versus the photon
energy $\omega$ for a single particle. The velocity v=0.99,
$E=100$ GeV, $m=14$ GeV, The field $F=100$ GeV$^2$, $T=0.1$
GeV$^{-1}$. } \label{Fig231}
\end{figure}
\begin{figure}
\centerline{\epsfig{figure=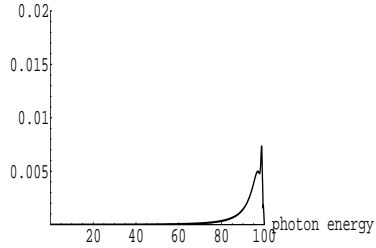,width=5cm,height=5cm,clip=}}
\caption{Spectral curve of a number of the radiated photons as a
function of $\omega$ for the very small time regime $T\le m/F$.
The parameters are the same as in the previous picture:E=100 GeV,
F=100 GeV$^2$,v=0.99. The transverse velocity in the c.m. frame of
the dipole is $v_{0t}=0.2$. } \label{Fig271}
\end{figure}
     \begin{figure}[htbp]
\centerline{\epsfig{figure=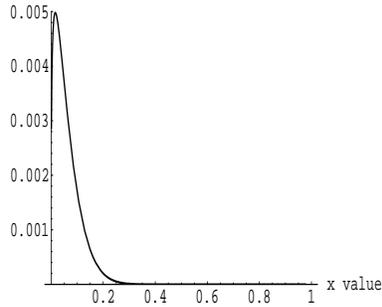,width=5cm,height=5cm,clip=}}
\caption{Spectral curve as a function of $x=\omega/E$ for a single
charged particle. We choose here F=100 GeV$^2$, E=100 GeV, v=0.8,
$\chi=0.04$. (All graphs for this and 6 figures below depict
$dE/(dTd\omega)$, normalized by $q^2/\sqrt{\pi}$.)} \label{Fig71}
\end{figure}
     \begin{figure}[htbp]
\centerline{\epsfig{figure=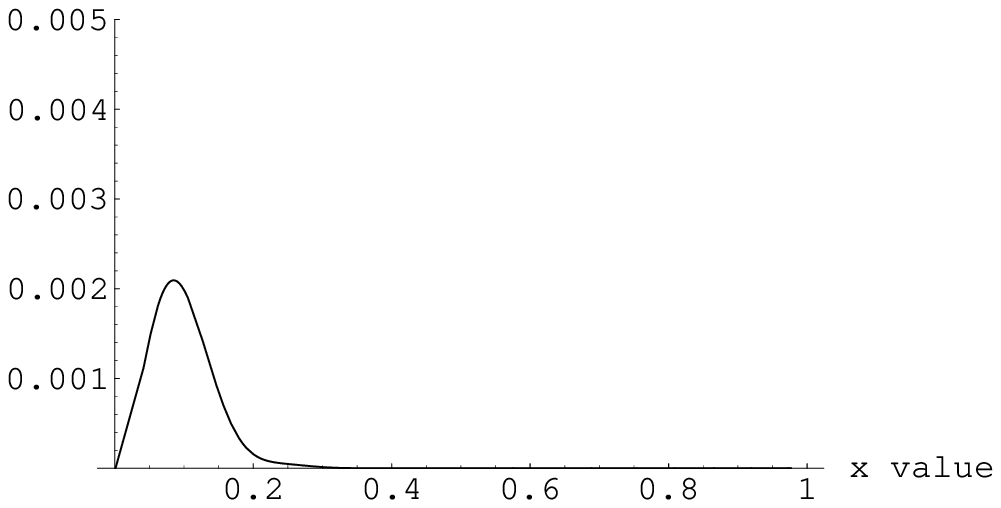,width=5cm,height=5cm,clip=}}
\caption{ Spectral curve as a function of $x=\omega/E$ for the
dipole in the small time regime. F,E,v are the same as for the
previous Figure, $\chi=0.04$, $T= m/F=25 $ GeV$^{-1}$.  }
\label{Fig81}
\end{figure}
     \begin{figure}[htbp]
\centerline{\epsfig{figure=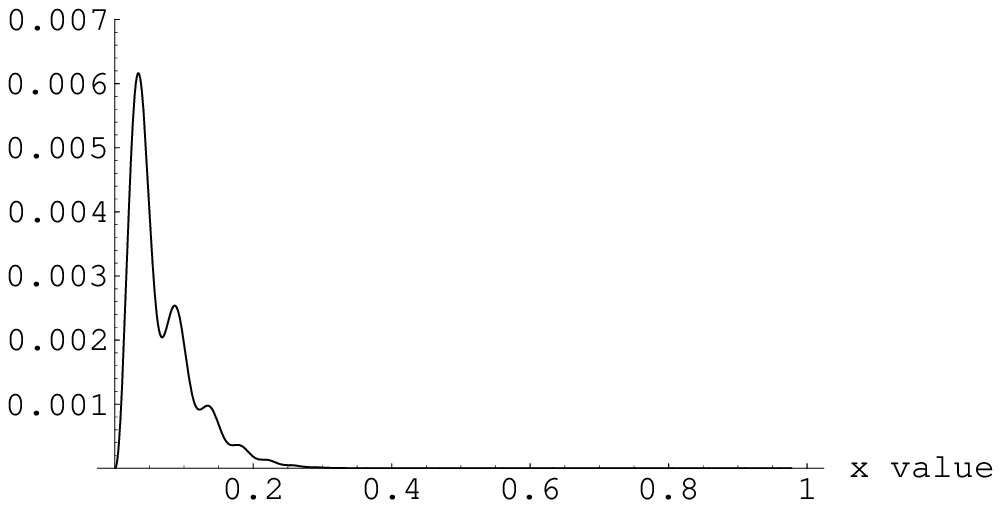,width=5cm,height=5cm,clip=}}
\caption{Spectral curve as a function of $x=\omega/E$ for the
dipole in the  small time regime. F,E,v are the same as for the
previous Figure, $\chi=0.04$, $T= E/F=100 $ GeV$^{-1}$.   }
\label{Fig91}
\end{figure}
    \begin{figure}[htbp]
\centerline{\epsfig{figure=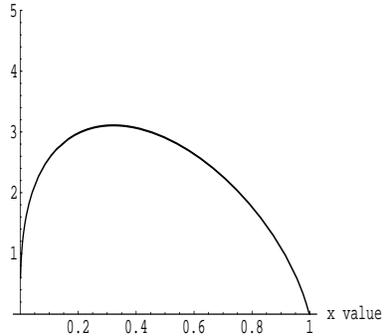,width=5cm,height=5cm,clip=}}
\caption{Spectral curve as a function of $x=\omega/E$ for a single
charged particle. We choose here F=100 GeV$^2$, E=100 GeV,
v=0.999, $\chi=111.6$. } \label{Fig101}
\end{figure}
     \begin{figure}[htbp]
\centerline{\epsfig{figure=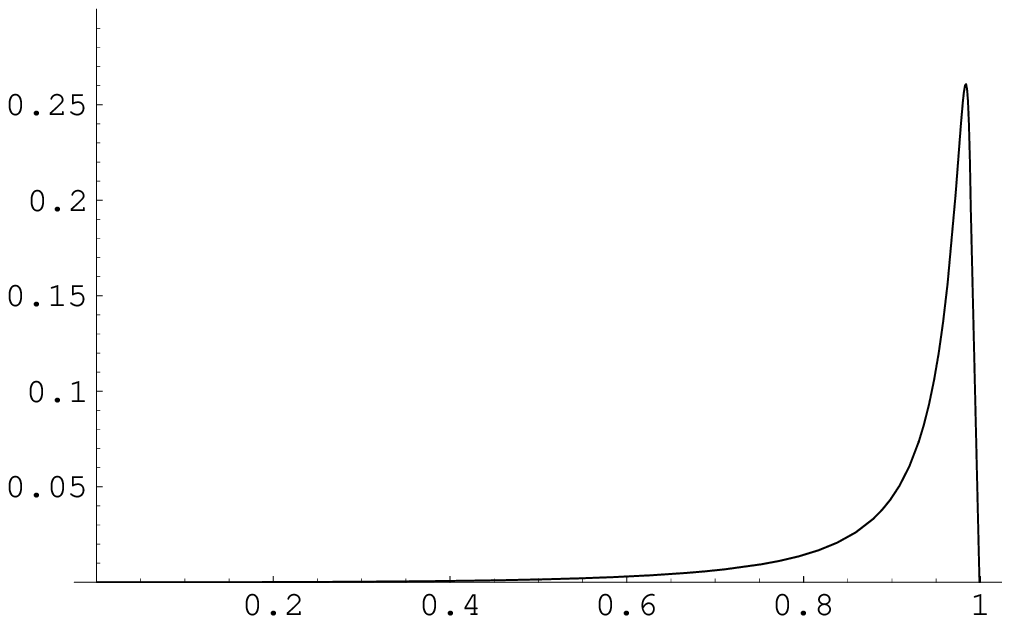,width=5cm,height=5cm,clip=}}
\caption{ Spectral curve as a function of $x=\omega/E$ for the
dipole in the small time regime. F,E,v are the same as for the
previous Figure, $\chi=111$, $T= m/F=0.045 $ GeV$^{-1}$.  }
\label{Fig111}
\end{figure}
     \begin{figure}[htbp]
\centerline{\epsfig{figure=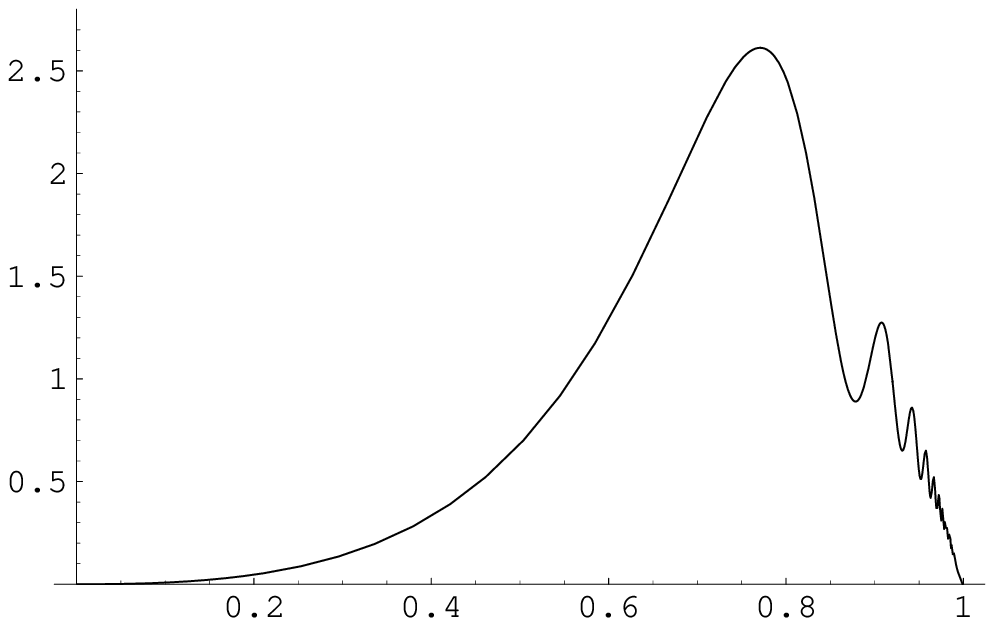,width=5cm,height=5cm,clip=}}
\caption{Spectral curve as a function of $x=\omega/E$ for the
dipole in the  small time regime. F,E,v are the same as for the
previous Figure, $\chi=111$, $T= (E/F^2)^{1/3}=0.21 $ GeV$^{-1}$.
 } \label{Fig121}
\end{figure}
     \begin{figure}[htbp]
\centerline{\epsfig{figure=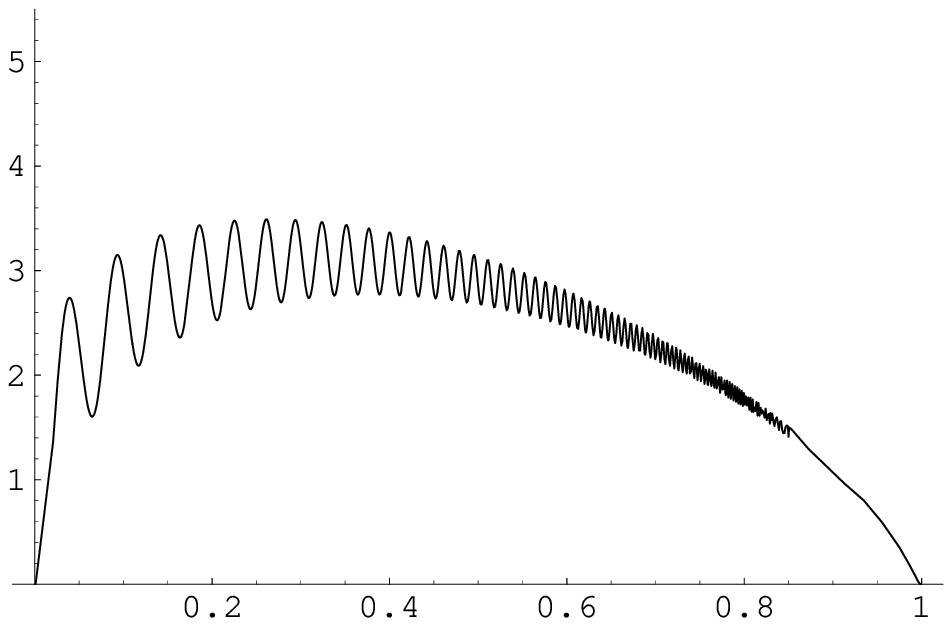,width=5cm,height=5cm,clip=}}
\caption{Spectral curve as a function of $x=\omega/E$ for the
dipole in the  small time regime. F,E,v are the same as for the
previous Figure, $\chi=111$, $T= E/F=100 $ GeV$^{-1}$.   }
\label{Fig131}
\end{figure}
     \begin{figure}[htbp]
\centerline{\epsfig{figure=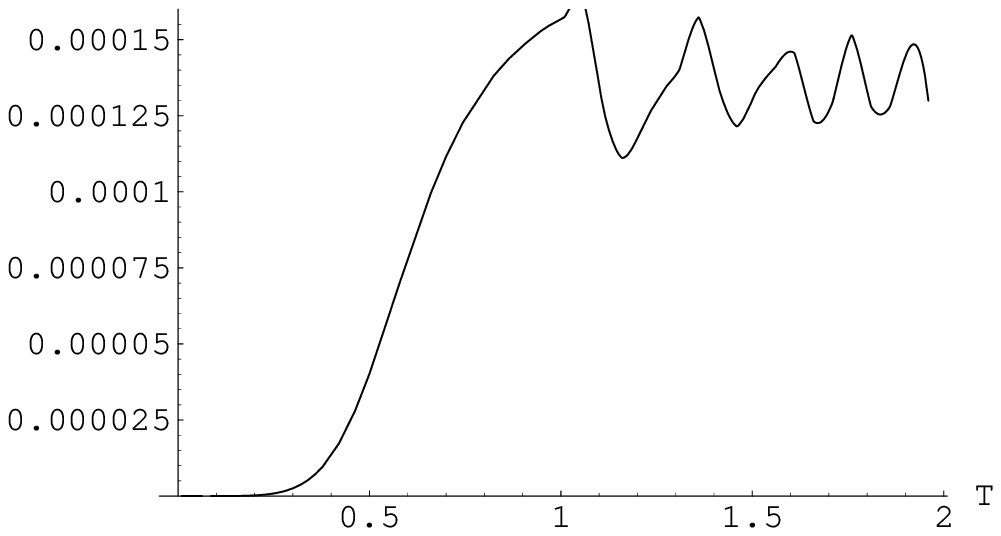,width=5cm,height=5cm,clip=}}
\caption{ Radiation Reaction of the dipole as a function of $T$
GeV$^{-1}$ for  small time regime $E/F\ge T\ge m/F$. For the
picture $\chi=0.04$, E=100 GeV, F=100 GeV$^2$. The radiation
reaction $dE/dT$ is normalized by $q^2/\sqrt{\pi}$. }
\label{Fig202}
\end{figure}     \begin{figure}[htbp]
\centerline{\epsfig{figure=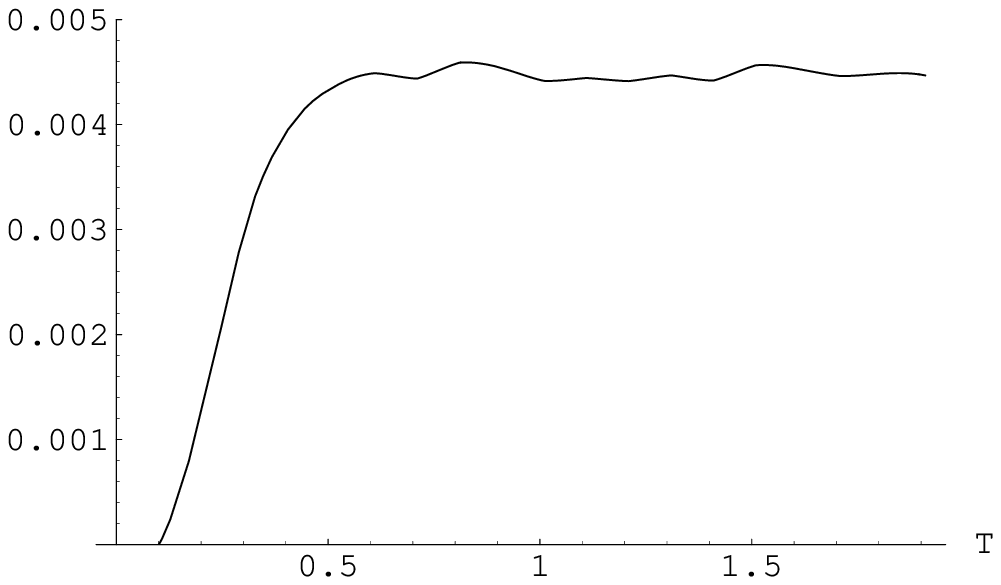,width=5cm,height=5cm,clip=}}
\caption{ Radiation Reaction of the dipole as a function of $T$
GeV$^{-1}$ for  small time regime $E/F\ge T\ge m/F$. For the
picture $\chi=111.4$, E=100 GeV, F=100 GeV$^2$, $T^*\sim 0.22$
GeV$^{-1}$. The radiation reaction $dE/dT$ is normalized by
$q^2/\sqrt{\pi}$. } \label{Fig203}
\end{figure}
\begin{figure}[htbp]
\centerline{\epsfig{figure=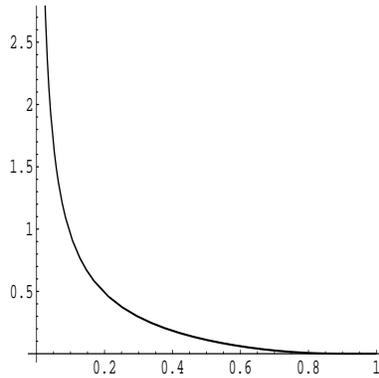,width=5cm,height=5cm,clip=}}
\caption{  Number of radiated photons in the  small dipole regime
as a function of frequency: no interference. (Normalized by
$q^2/\sqrt{\pi}$ .)} \label{Fig11}
\end{figure}
\begin{figure}[htbp]
\centerline{\epsfig{figure=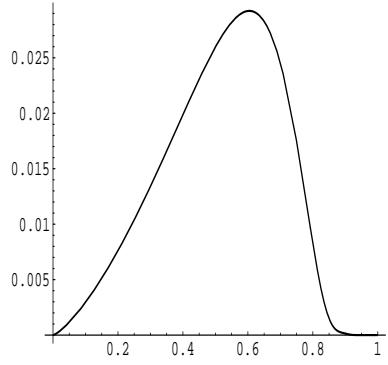,width=5cm,height=5cm,clip=}}
\caption{Number of radiated photons in the small dipole regime.
Interference is taken into account.(Normalized by
$q^2/\sqrt{\pi}$).} \label{Fig12}
\end{figure}
\begin{figure}[htbp]
\centerline{\epsfig{figure=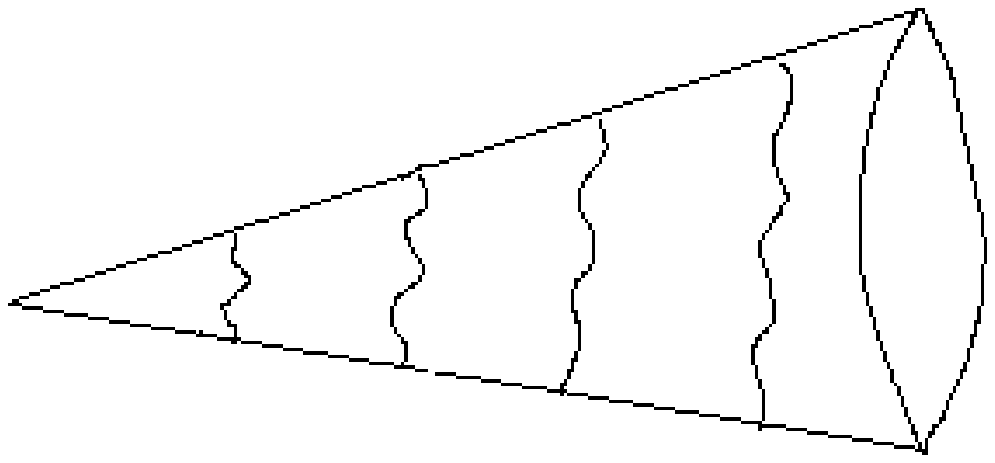,width=5cm,height=5cm,clip=}}
\caption{The graphs with the noncoulombic photon exchange that
lead to quantum diffusion.  } \label{Fig16}
\end{figure}


\newpage\begin{references}
\bibitem{LL} L. Landau and E. Lifshits, Field Theory, Pergamon
Press,1978.
\bibitem{TM} F. Rohlich, Relativistic particle Electrodynamics:
How its Problem Got Resolved. Vikram A. Sarabhai Lectures, January
1992.
\bibitem{R} F.Rohlich, Classical Charged Particles, Addison-Wesley Publishing Co.,
 Redwood City,CA 1965 and 1990 ; J.D. Jackson, Classical Electrodynamics,
Wiley, New York (1975).
\bibitem{LL1} V.B. Berestetskii, E.M. Lifshits, L.P. Pitaevski, Quantum 
Electrodynamics, Pergamon Press, 1982.
\bibitem{F} E.S. Fradkin, D. Gitman, S.M. Shvartsman, Quantum
Electrodynamics with unstable vacuum, Springer-Verlag,1991 and
references therein.
\bibitem{NR} A.I. Nikishov, ZhETP, 59 (1970) 1262;
Problems of Intense External Field in Quantum Electrodynamics, in
Quantum Electrodynamics of Phenomena in Intense Field, Proc. P.N.
Lebedev Institute, vol. 111 (Nauka, Moscow, 1979),pp. 153; A.I.
Nikishov and V.I. Ritus ZhetF, 46 (1964) 776,1778; 47 (1964) 1130.
\bibitem{Baier} V. N. Baier, V.N. Katkov, V.S. Fadin, Radiation  of
Relativistic Electron, (quit, Moscow, 1973).
\bibitem{ST} A.A. Sokolov, I.M. Ternov, The Relativistic Electron (Nauka,
Moscow, 1974) and references therein.
\bibitem{AS1} A. I. Akhiezer and N. F. Shulga, High-Energy Electrodynamics
in Matter, Gordon and Breach Publishers, 1996.
\bibitem{AS2} A.I. Akhiezer and N. F. Shulga, Phys. Reports, 234 (1993)
296.
\bibitem{Schifffe}50) R. Baier, Yuri L. Dokshitzer, Alfred H. Mueller, S.
    Peigne, D. Schiff,
    Nucl.Phys., B484 (1997) 265-282.
\bibitem{Feinberg} E.L. Feinberg, Sov. Phys. JETP 23 (1966) 132;
in Problems of Theoretical Physics. A Memorial volume to I.E.
Tamm, Nauka, Moscow (1972).
\bibitem{Perkins} D. Perkins , Phil. Mag., 46 (1955) 1146;
J. Iwadare, Phil. Mag., 3 (1958) 680.
\bibitem{Chudakov} A.E. Chudakov, Izv. Akad. Nauk, USSR, (Ser.
Fiz.), 19 (1955) 650.
\bibitem{Iekutieli} G. Iekutieli, Nuovo Cim., 5 (1957) 1381; I.
Mito, H. Ezawa, Progr. Theor. Phys., 18 (1957) 437; G.H.
Burkhardt, Nuovo Cim., 9 (375) 1958.
\bibitem{Farrar}G. R. Farrar , H. Liu, L. Frankfurt and M. Strikman,
 Phys. Rev. Lett.,61 (1988) 686; 62 (1989)
387.
\bibitem{FS1} L. Frankfurt and M. Strikman, Phys. Rept., 160
(1988) 235.
\bibitem{M} A.H. Mueller, Nucl. Phys., B415 (1994) 373; B437 (1995)
107.
\bibitem{FS} L. Frankfurt,J. Miller, M. Strikman, Annual review on
Particle and Nuclear Physics, 44 (1994) 501.
\bibitem{CFS} J. S. Collinz, L. Frankfurt, M. Strikman, Phys. Rev.
, D56 (1997) 2982.
\bibitem{Imri} A.B. Migdal, Fermions and bosons in strong fields
(in russian), Nauka, Moscow, 1978;Y. Imri, 
Introduction to 
mesoscopic physics, Oxford, Oxford University press, 1997.
\bibitem{AS}M. Abramowitz and I. Stegun eds., Handbook on mathematical
functions, NBS, 1964
\bibitem{BE} H. Beitmen and A. Erdelyi, Higher Trancsedental Functions,
vol. 2, McGraw-Hill Book Company, 1955.
\bibitem{WW} E.T. Whittaker and G.N. Watson, A course of Modern Analysis,
Cambridge University Press, 1927.
\bibitem{W} G.N. Watson, Bessel Functions, Cambridge University Press,
2nd edition, 1944.
\bibitem{DP} A.P. Prudnikov, Yu. A. Brychkov, O.I. Marichev,
Integrals and Series, v. 2:Special Functions. Gordon and Breach
Science Publishers, 1992 (3d printing).
\bibitem{Doc}Yu. L. Dokshitser et al,Basics of  Perturbative QCD.
Edition Frontieres, Gif-Sur-Yvette Cedex, France, 1991.
\bibitem{M1}A.H. Mueller, Nucl. Phys., B307 (1988) 34, B317 (1989)
573.
\bibitem{LP}L. Landau, R. Pierels, Zs. phys., 69 (1931) 56.

\end{references}
\end{document}